\documentclass{emulateapj}
\usepackage{lscape}

\newcommand\etal{et al.\ }
\newcommand\eg{{\it e.g.},\ }
\newcommand\ie{{\it i.e.},\ }
\newcommand\rmag{\ifmmode r_{625}\else$r_{625}$\fi}
\newcommand\imag{\ifmmode i_{775}\else$i_{775}$\fi}
\newcommand\zmag{\ifmmode z_{850}\else$z_{850}$\fi}
\newcommand\Vmag{\ifmmode V_{606}\else$V_{606}$\fi}
\newcommand\Imag{\ifmmode I_{814}\else$I_{814}$\fi}

\newcommand\feso{\mathrel{f_{\rm E{+}S0}}}
\newcommand\fso{\mathrel{f_{\rm S0}}}
\newcommand\fsp{\mathrel{f_{\rm Sp{+}Irr}}}
\newcommand\fe{\mathrel{f_{\rm E}}}
\newcommand\lx{\mathrel{L_{\rm x}}}
\newcommand\lxbol{\mathrel{L_{\rm x,Bol}}}
\newcommand\tempx{\mathrel{T_{\rm x}}}
\newcommand\rtwoh{\ifmmode {\rm r}_{200}\else r$_{200}$\fi}
\newcommand\ls{\mathrel{\hbox{\rlap{\hbox{\lower4pt\hbox{$\sim$}}}\hbox{$<$}}}}
\newcommand\gs{\mathrel{\hbox{\rlap{\hbox{\lower4pt\hbox{$\sim$}}}\hbox{$>$}}}}

\def\simgreat{\ifmmode{\mathrel{\mathpalette\@versim>}}
    \else{$\mathrel{\mathpalette\@versim>}$}\fi}
\def\simless{\ifmmode{\mathrel{\mathpalette\@versim<}}
    \else{$\mathrel{\mathpalette\@versim<}$}\fi}

\newcounter{thefigs}

\newcounter{thetabs}

\newcommand{\hGpc}{\ifmmode{h^{-1}{\rm Gpc}}\;\else${h^{-1}}${\rm Gpc}\fi}

\def\simless{\mathbin{\lower 3pt\hbox
	{$\,\rlap{\raise 5pt\hbox{$\char'074$}}\mathchar"7218\,$}}} 
\def\simgreat{\mathbin{\lower 3pt\hbox
	{$\,\rlap{\raise 5pt\hbox{$\char'076$}}\mathchar"7218\,$}}} 

\shortauthors{Postman \etal}
\shorttitle{The Morphology--Density Relation in $z \sim 1$ Clusters}
\slugcomment{Accepted for publication in the {\it Astrophysical Journal}} 

\begin{document}

\title{The Morphology--density Relation in $z \sim 1$ Clusters}
\author{
M. Postman\altaffilmark{1},
M. Franx\altaffilmark{2},
N.J.G. Cross\altaffilmark{3},
B. Holden\altaffilmark{4},
H.C. Ford\altaffilmark{3},
G.D. Illingworth\altaffilmark{4},
T. Goto\altaffilmark{3},
R. Demarco\altaffilmark{3},
P. Rosati\altaffilmark{5}, 
J.P. Blakeslee\altaffilmark{3},
K.-V. Tran\altaffilmark{6},
N. Ben\'{\i}tez\altaffilmark{3},
M. Clampin\altaffilmark{7},
G.F. Hartig\altaffilmark{1},
N. Homeier\altaffilmark{3},
D.R. Ardila\altaffilmark{3},
F. Bartko\altaffilmark{8}, 
R.J. Bouwens\altaffilmark{4},
L.D. Bradley\altaffilmark{3},
T.J. Broadhurst\altaffilmark{9},
R.A. Brown\altaffilmark{1},
C.J. Burrows\altaffilmark{10},
E.S. Cheng\altaffilmark{11},
P.D. Feldman\altaffilmark{3},
D.A. Golimowski\altaffilmark{3},
C. Gronwall\altaffilmark{12},
L. Infante\altaffilmark{13},
R.A. Kimble\altaffilmark{7},
J.E. Krist\altaffilmark{1},
M.P. Lesser\altaffilmark{14},
A.R. Martel\altaffilmark{3},
S. Mei\altaffilmark{3},
F. Menanteau\altaffilmark{3},
G.R. Meurer\altaffilmark{3},
G.K. Miley\altaffilmark{2},
V. Motta\altaffilmark{13},
M. Sirianni\altaffilmark{1}, 
W.B. Sparks\altaffilmark{1}, 
H.D. Tran\altaffilmark{15}, 
Z.I. Tsvetanov\altaffilmark{3},   
R.L. White\altaffilmark{1}
\& W. Zheng\altaffilmark{3}
}
\altaffiltext{1}{Space Telescope Science Institute, 3700 San Martin Drive, Baltimore, MD 21218.}
\altaffiltext{2}{Leiden Observatory, Postbus 9513, 2300 RA Leiden, Netherlands.}
\altaffiltext{3}{Department of Physics and Astronomy, Johns Hopkins University, 3400 N. Charles Street, Baltimore, MD 21218.}
\altaffiltext{4}{UCO/Lick Observatory, University of California, Santa Cruz, CA 95064.}
\altaffiltext{5}{European Southern Observatory, Karl-Schwarzschild-Strasse 2, D-85748 Garching, Germany.}
\altaffiltext{6}{Institute for Astronomy, ETH Z\"urich, Scheuchzerstrasse 7, CH-8093 Zürich, Switzerland.}
\altaffiltext{7}{NASA Goddard Space Flight Center, Code 681, Greenbelt, MD 20771.}
\altaffiltext{8}{Bartko Science \& Technology, 14520 Akron Street,Brighton, CO 80602.}
\altaffiltext{9}{Racah Institute of Physics, The Hebrew University, Jerusalem, Israel 91904.}
\altaffiltext{10}{MetaJiva Scientific.}
\altaffiltext{11}{Conceptual Analytics, LLC, 8209 Woburn Abbey Road, Glenn Dale, MD 20769}
\altaffiltext{12}{Department of Astronomy and Astrophysics, The Pennsylvania State University, 525 Davey Lab, 
                  University Park, PA 16802.}
\altaffiltext{13}{Departmento de Astronom\'{\i}a y Astrof\'{\i}sica, Pontificia Universidad Cat\'{\o}lica de Chile, 
                  Casilla 306, Santiago 22, Chile.}
\altaffiltext{14}{Steward Observatory, University of Arizona, Tucson, AZ 85721.}
\altaffiltext{15}{W. M. Keck Observatory, 65-1120 Mamalahoa Hwy., Kamuela, HI 96743}

\begin{abstract}
We measure the morphology--density relation (MDR) and morphology-radius relation (MRR) 
for galaxies in seven $z \sim 1$ clusters that have been observed with the Advanced Camera for Surveys
on board the Hubble Space Telescope. Simulations and independent comparisons of our
visually derived morphologies indicate that ACS allows one to distinguish 
between E, S0, and spiral morphologies down to $\zmag = 24$, corresponding
to $L/L^* = 0.21$ and $0.30$ at $z = 0.83$ and $z=1.24$, respectively. 
We adopt density and radius estimation methods
that match those used at lower redshift in order to study the evolution of the MDR and MRR. 
We detect a change in the MDR
between $0.8 < z < 1.2$ and that observed at $z \sim 0$, consistent with
recent work -- specifically, the growth in the bulge-dominated galaxy 
fraction, $\feso$, with increasing density proceeds less rapidly
at $z \sim 1$ than it does at $z \sim 0$. At $z \sim 1$ and $\Sigma \ge 500$ galaxies Mpc$^{-2}$, we find
$<\feso> = 0.72\pm0.10$. At $z \sim 0$, an E+S0 population fraction of this magnitude occurs at densities about
5 times smaller. The evolution in the MDR is confined to densities $\Sigma \simgreat 40$
galaxies Mpc$^{-2}$ and appears to be 
primarily due to a deficit of S0 galaxies and an excess of Spiral+Irr galaxies relative
to the local galaxy population. The $\fe$ -- density relation 
exhibits no significant evolution between $z = 1$ and $z = 0$.
We find mild evidence to suggest that the MDR is dependent on
the bolometric X-ray luminosity of the intracluster medium.  
Implications for the evolution of the disk galaxy population in dense
regions are discussed in the context of these observations.
\end{abstract}


\section{Introduction}\label{intro}

The study of the origin and evolution of the morphological distribution of galaxies in different
environments can reveal important 
information about internal galactic stellar and gas dynamics, the state of star formation 
activity as a function of time, and constrain the relative significance of the
effects of environmental processes versus conditions at the epoch of their formation
in establishing galactic structure. In standard hierarchical clustering 
models, galaxies in high density regions of the Universe, such as in the central
regions of galaxy clusters, will collapse earlier and may evolve more rapidly than 
galaxies in low density regions (Kauffmann 1995; Benson \etal 2001, Heavens \etal 2004). In addition, 
galaxies in dense environments are subject to a variety of external stresses, which are,
in general, not conducive to the maintenance of spiral structure.
These processes include ram-pressure stripping of gas (Gunn \& Gott 1972; Farouki
\& Shapiro 1980; Kent 1981; Fujita \& Nagashima 1999; Abadi, Moore \& Bower 1999; 
Quilis, Moore \& Bower 2000), galaxy harassment via high speed impulsive
encounters (Moore \etal 1996, 1999; Fujita 1998), cluster tidal forces (Byrd \& 
Valtonen 1990; Valluri 1993; Fujita 1998) which distort galaxies as they come close 
to the center, interaction/merging (Icke 1985; Lavery \& Henry 1988, 
Mamon 1992; Makino \& Hut 1997; Bekki 1998), and 
removal and consumption of the gas due to the cluster environment 
(Larson, Tinsley \& Caldwell 1980; Bekki \etal 2002). 

Two of key relationships that must be understood in the context of the above processes
are the relative population fraction of the different
morphological classes as functions of the local galaxy density and 
their location within the local gravitational potential well. 
The morphology -- density relation (hereafter MDR) and the morphology -- radius relation (hereafter MRR) have 
been well studied at low-$z$ (Dressler 1980 - hereafter D80;
Postman \& Geller 1984 - hereafter PG84; Whitmore \& Gilmore 1991; Goto \etal 2003a) 
and quantify many long-standing observations showing a preference
for spheroidal systems to reside in dense regions (or perhaps better stated as
a significant lack of spiral galaxies in dense regions). A full understanding of how
such a cosmic arrangement came to be requires measuring the evolution of the MDR and MRR.
Such an evolutionary study is only possible using the high angular resolution provided by the Hubble
Space Telescope (HST).
Several pioneering works have now shed light on this evolution (Dressler \etal 1997 -- hereafter D97; Fasano \etal 2000;
Treu \etal 2003; Smith \etal 2004 -- hereafter Sm04). D97 and Fasano \etal (2000) find a significant decline in the
fraction of lenticular galaxies ($\fso$) when one looks back from the present epoch to an epoch $4 - 5$ Gyr ago.  
The results presented by Treu \etal 2003 and Sm04 are perhaps the most
enlightening - they find a smaller increase in the bulge-dominated galaxy (E+S0) fraction ($\feso$) with increasing
density at $z \simgreat 0.4$ than is seen at $z < 0.1$ but also find comparable $\feso$ values for low-density
regions ($\Sigma < 10$ galaxies Mpc$^{-2}$) 
at $z \simgreat 0.4$ and the current epoch. Sm04 propose a simple model to explain these
observations in which high density regions at $z \sim 1$ would largely be comprised of elliptical
galaxies with only a trace of lenticulars (\eg $0 \le \fso < 0.1$). They consider various processes to
transform spiral galaxies into lenticular galaxies in order to increase $\fso$
with time to match the observed morphological population fractions
at $z \sim 0.5$. However, the Sm04 $z \sim 1$ $\fso$ 
measurement was inferred from $\feso$ rather than measured directly as Sm04 chose (perhaps wisely) 
not to attempt to distinguish between S0 and E galaxies from the WFPC2 data used in their study.

The deployment of the Advanced Camera for Surveys (ACS; Ford \etal 2003) on the HST
has provided us with an opportunity to greatly
expand our understanding of the physics behind the morphological evolution of
galaxies in a wide variety of environments. The higher sensitivity and better angular
sampling of the Wide Field Camera (WFC) on ACS relative to WFPC2 enables
the acquisition of high S/N morphological information for sub-$L^{*}$ galaxies
over projected areas of up to 10 Mpc$^{2}$ in $z \sim 1$ clusters, in a
modest allocation of telescope time. This is a significant advantage over prior capabilities
and enables us to sample $>3$ decades in local galaxy density using the same homogeneous
data samples. 

As part of an extensive program to study the formation and evolution of
clusters and their galaxy populations, the ACS Investigation Definition Team (IDT) has implemented a
128 orbit program to observe 7 distant clusters in the redshift range
$0.83 \le z \le 1.27$. In this paper, we present new constraints on the form
and evolution of both the MDR and the MRR in these
clusters and their surroundings
based on morphologies determined from the ACS/WFC imagery coupled with
extensive spectroscopic data and X-ray observations.
This paper is organized as follows: section~\ref{data} contains a brief summary
of the space and ground-based observations used in this study, section~\ref{morph}
presents a detailed discussion of our morphological classification procedure and
an assessment of the reliability of these classifications, section~\ref{density}
presents the methods used to estimate the local projected density, and 
section~\ref{mdrresults} presents our derived MDR and MRR. An assessment of
the implications of these results is given in section~\ref{discuss} and a summary
of the essential results is provided in section~\ref{conclude}. Two appendices discuss
details associated with the computation population
fractions that are suitably corrected for contamination and incompleteness, the robustness
of our density estimators, and the validity of using composite samples to enhance the
signal-to-noise ratio in the derived MDR and MRR.
We adopt $H_o = 70$ km s$^{-1}$ Mpc$^{-1}$, $\Omega_m = 0.3$, and
$\Omega_{\Lambda} = 0.7$ for the computation of all intrinsic quantities
unless specifically indicated otherwise.

\section{Observations}\label{data}

The clusters included in this study, along with a summary of the ACS observations, 
are listed in Table~\ref{tab_obssum}.
The average cluster redshift, based on all available spectroscopically confirmed
cluster members, is given in column 2 of this table. The number of redshifts acquired
for galaxies in the region of each cluster (both members and non-members and
including galaxies outside the ACS mosaic boundaries) is listed in column 3.
Column 4 lists the number of spectroscopically confirmed cluster members that
also lie within the ACS mosaic boundaries. The details of the HST ACS observations
are given in columns 5 -- 7.
The sample selection process was limited by the small number of 
spectroscopically confirmed clusters at $z > 0.8$. However, we were able to
include clusters with a range of X-ray luminosities, 
from $\lxbol < 10^{44}$ erg s$^{-1}$
to $\lxbol = 1.9 \times 10^{45}$ erg s$^{-1}$.  Table~\ref{tab_xray} is a compilation of
the derived X-ray properties and velocity dispersions of these clusters. Two of the seven
clusters (the two at R.A. = 16 hr) are optically selected systems, 
the rest are X-ray selected although RXJ0849+4452 is a binary
cluster system in which the less massive component (CL J0848+4453)
was IR-selected (Stanford \etal 1997).

\subsection{ACS Observations}\label{space}

We used the WFC on the ACS to image each cluster. 
For MS1054-0321, RXJ0152-1357, RDCS1252-2927, and RXJ0849+4452 
multiple pointings were used to construct
larger mosaics covering $\sim35$ square arcminutes. For the first three
of these clusters, the pointings form a 2 $\times$ 2 pattern with all
4 pointings overlapping the central 1 arcminute region of each
cluster - hence the exposure time in the central regions of these systems is
4 times as long as the values given in Table~\ref{tab_obssum}. For RXJ0849+4452, we used
a 3 $\times$ 1 pattern in order to obtain images of both components of this
binary cluster system. All the remaining targets were observed using
a single WFC pointing centered on the cluster. 

The filters are chosen to approximately straddle the rest-frame 4000\AA\ break in
order to facilitate the identification of bulge-dominated galaxies in the red sequence of
the clusters. This sequence, which is populated mostly by elliptical and lenticular galaxies with a strong 4000\AA\ break,
is well separated from the CM relation for late-type cluster galaxies as well as that for most field galaxies.
In all cases, we have at least one filter that samples part of the
rest-frame $B-$band. We use the ACS images taken in those filters to perform
our morphological classifications so that we can readily
compare our results with morphological information obtained at lower redshifts (e.g., Fabricant, Franx, \&
van Dokkum 2000).
The filters used are explicitly listed in Table~\ref{tab_obssum}. Hereafter, we will use
\Vmag\ to denote the F606W bandpass, \rmag\ to denote F625W,
\imag\ to denote F775W, \Imag\ to denote F814W, and \zmag\ to denote F850LP. 

\subsection{Object Photometry and Classification}\label{apsis}

Object detection and analysis is performed using the ACS IDT Pipeline ({\it a.k.a.} APSIS;
Blakeslee \etal 2003a). APSIS photometry is on the AB system and is corrected for Galactic
extinction using the Schlegel \etal (1998) $100\mu$ map. Object detection and
star-galaxy discrimination is done using the dual image mode in SExtractor (Bertin \& Arnouts 1996). 
The detection image is an inverse variance weighted combination of the ACS exposures
from all available passbands. The inverse variance weighting preserves information
about the structural characteristics of the galaxies (see Ben\'{i}tez \etal\ 2004 for details). In this paper, we count
as galaxies all objects with SExtractor stellarity parameter {\tt CLASS\_STAR} $\le$ 0.50.
The automated image structure analysis of detected objects in our ACS data 
includes the determination of the luminosity-weighted
moments, the ellipticity, and the 180$^{\circ}$ rotational asymmetry 
and image concentration parameters (\eg Abraham \etal 1994; Conselice \etal 2000).
All magnitudes cited in this study are based on the SExtractor {\tt MAG\_AUTO} magnitude
as it provides a reasonable estimate of the total flux.

\subsection{Spectroscopic Observations and Photometric Redshifts}\label{specobs}

Spectroscopic redshifts have been obtained for the clusters in our survey, by us and others, using 
multiobject spectrographs on the Keck, VLT, or the Magellan observatories. The total number 
of redshifts and the number of confirmed cluster members within each ACS mosaic are listed in 
Table~\ref{tab_obssum}. The publications containing some or all of the 
redshift data include 
Tran \etal (2005) for MS1054-0321, 
Demarco \etal (2004) for RXJ0152-1357, 
Postman \etal (1998a, 2001) and Gal \& Lubin (2004) for the CL1604+43 system, 
Stanford \etal (2002) for RDCS0910+5422, 
Demarco \etal (2005) for RDCS1252-2927, and 
Rosati \etal (1999) for RXJ0849+4452. 
The target selection criteria for the redshift surveys 
of MS1054-0321 and the CL1604+43 clusters were based on a single red flux limit. The target selection 
for RXJ0152-1357 include a color-selection criterion (see Demarco \etal 2004). The spectra are of moderate 
resolution (R $\sim 500 - 1200$) and most have sufficient S/N to measure the prominent spectral
features (\eg [OII] line widths).

\begin{figure*}
\plotone{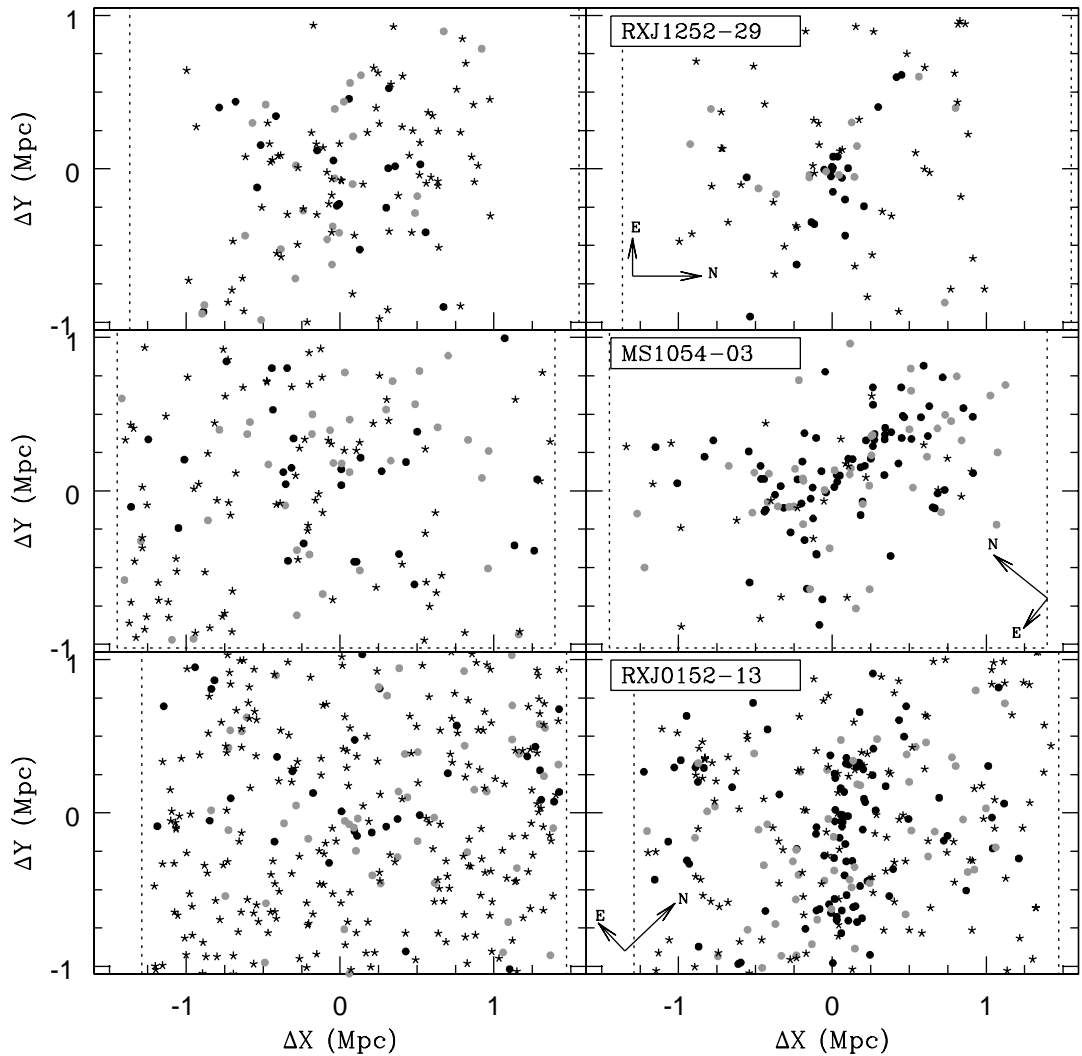}
\caption[]{
The projected distribution of galaxies with photometric redshifts that lie within $\pm2\sigma_{ph}$ 
of the mean cluster redshift are shown in the right-hand panels for RXJ0152-1357, MS1054-0321, and 
RDCS1252-2927. The distribution of galaxies with photometric redshifts in 
the range $2\sigma_{ph} < |\overline z_{\rm CL} - z_{ph}| / (1 + z_{\rm CL}) < 6\sigma_{ph}$ 
are shown in the left-hand panels.  Different symbols denote different morphological 
classifications: black dots are ellipticals, grey dots are S0's, and stars are Sp+Irr. 
The dashed lines denote the boundaries of the ACS mosaics. The RDCS1252-2927 photo-z's are
available over less area than the full ACS mosaic because they rely on NIR photometry that 
covers a smaller region.  The overdensities associated with the clusters
are easily seen in the right-hand panels and are dominated by E and S0 galaxies.}
\label{fig_bpzmap}
\end{figure*}

Our photometric redshifts are derived using the Bayesian method ({\it a.k.a.} BPZ) described 
in Ben\'{i}tez (2000) and are based on a minimum of 3 passbands, including all available ACS 
photometry. We have reliable photo-z's for RXJ0152-1357 ($z = 0.837$), MS1054-0321 ($z=0.831$), 
and RDCS1252-2927 ($z=1.235$). Photo-z's for RXJ0152-1357 are based on the ACS \rmag, \imag, and
\zmag\ photometry. Photo-z's for MS1054-0321 are based on the ACS \Vmag, \imag, and \zmag\ photometry.
Photo-z's for RDCS1252-2927 are based on ACS \imag, \zmag\ photometry and ground-based $BVJK$ photometry
(Toft \etal 2004). 
For the two $z=0.83$ clusters, the rms scatter in 
$(z_{spec} - z_{ph}) / (1 + z_{spec})$, $\ \sigma_{ph}$, is 0.05. For RDCS1252-2927,
$\sigma_{ph}\ $ is 0.10. Only objects that have a BPZ confidence level of 0.90 or greater are selected for analysis.
Figure~\ref{fig_bpzmap} shows the distribution of galaxies
with photometric redshifts within $2 \sigma_{ph}$ of the 
mean cluster redshift and galaxies with 
$2\sigma_{ph} < |\overline z_{\rm CL} - z_{ph}| / (1 + z_{\rm CL}) < 6\sigma_{ph}$. 
The cluster overdensity is clearly seen only when we select galaxies close to the actual mean
cluster redshift, indicating our photo-z's are useful in significantly suppressing 
fore/background contamination and isolating most of the actual cluster members.

\section{Morphological Classification}\label{morph}

\begin{figure*}
\plotone{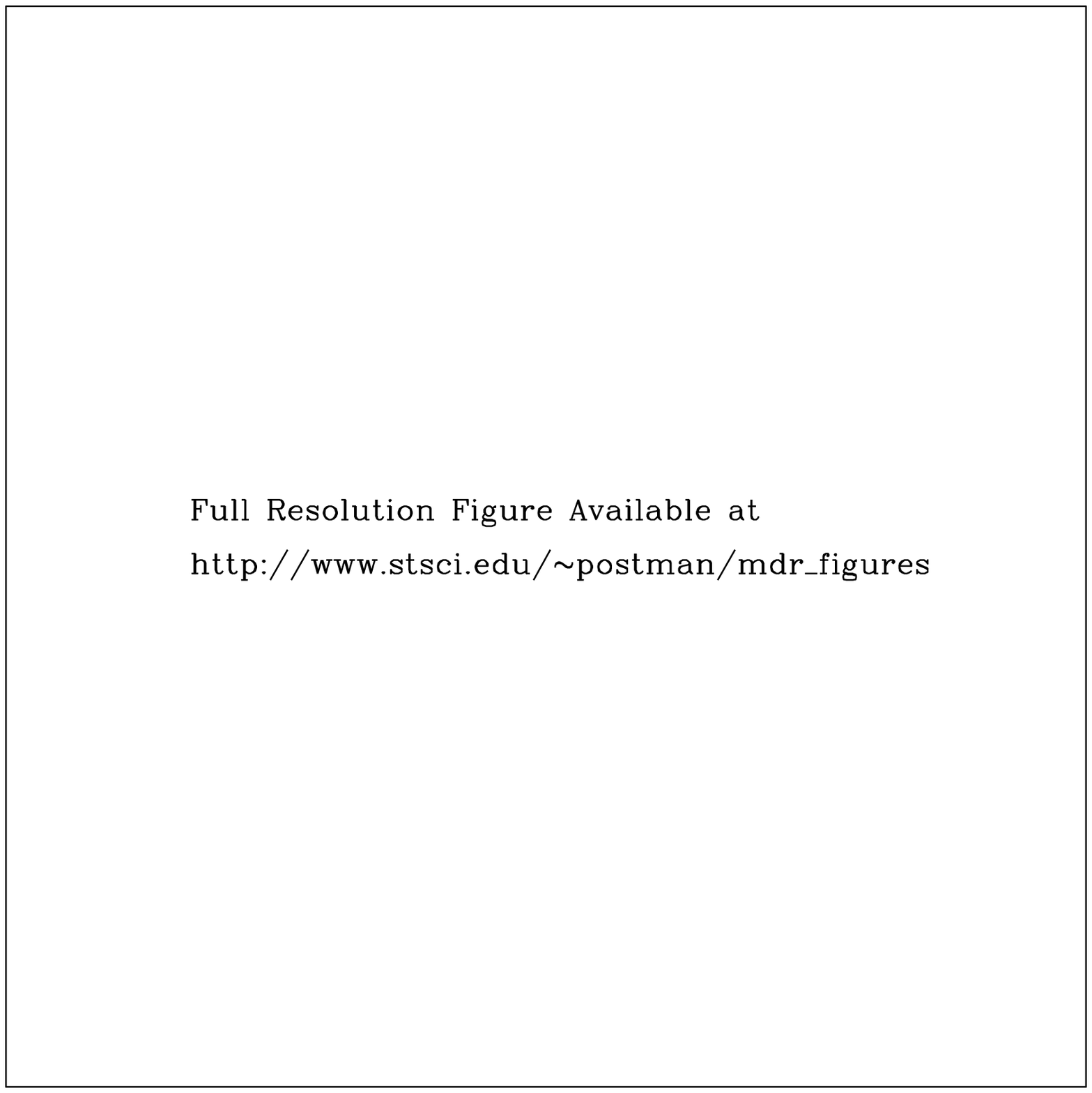}
\caption[]{
Color postage stamp cutouts of the brightest 49 spectroscopically confirmed members
of CL1604+4321 ($z = 0.92$) and CL1604+4304 ($z=0.90$) 
and their corresponding visually derived morphological
classifications. The ``Pos Disk" classification stands for ``possible disk" galaxy.
It is given to objects that appear to have a disk structure but the precise nature
of that disk could not established. Galaxies with the ``Pos Disk" classification
are counted as Spirals in our derivation of the MDR. 
The first 29 cells show galaxies from CL1604+4321 and the rest contain galaxies 
from CL1604+4304. The galaxies from each cluster are shown in increasing apparent
magnitude order and span the range $20.91 \le \imag\ \le 23.50$. Each cutout subtends
a $6.4 \times 6.4$ arcsecond area.
}
\label{fig_cl1604morfs}
\end{figure*}

\begin{figure*}
\plotone{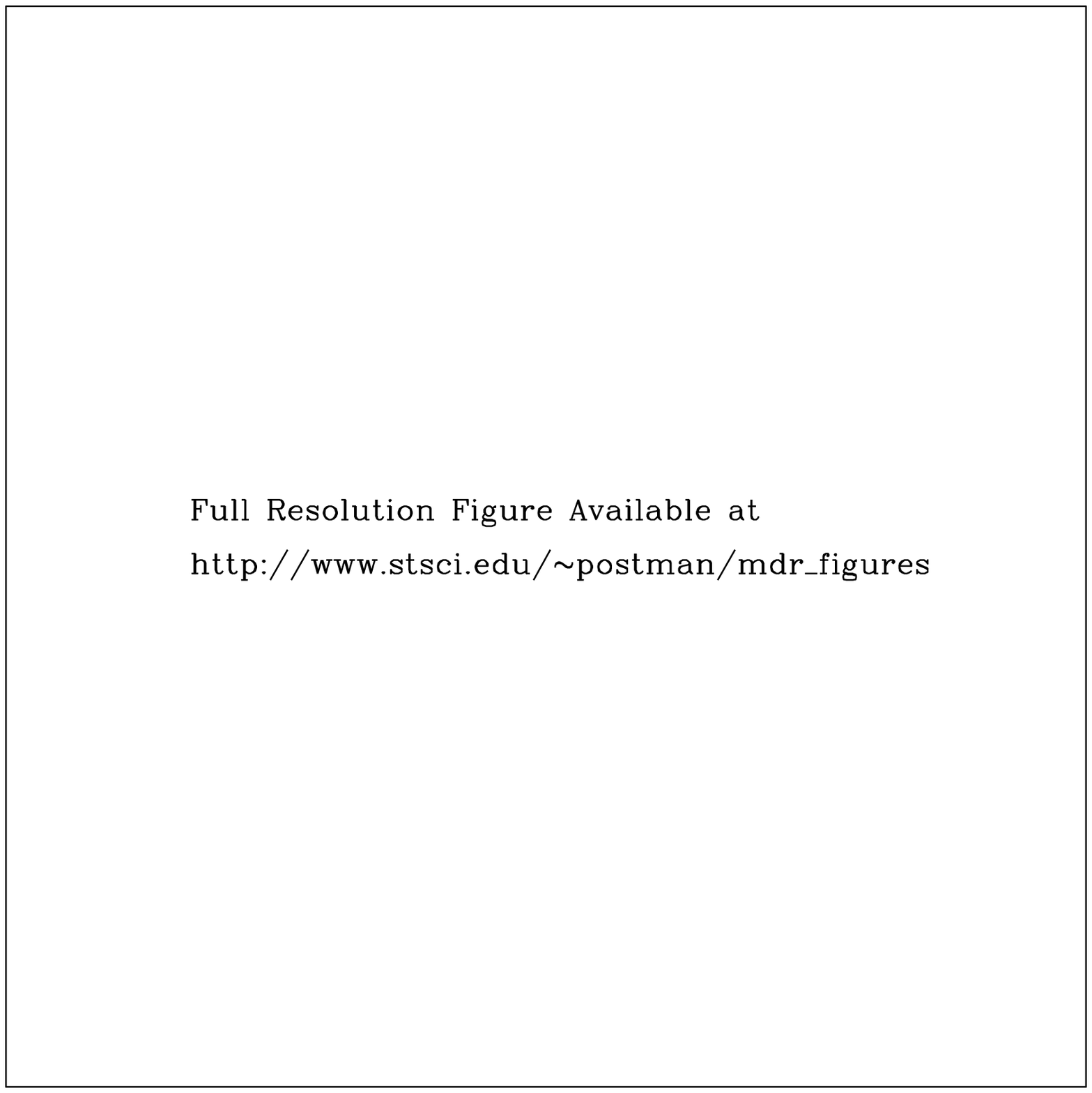}
\caption[]{
Color postage stamp cutouts of the brightest 49 spectroscopically confirmed members
of MS1054-0321 ($z = 0.831$).
The galaxies are displayed in increasing apparent
magnitude order and span the range $20.14 \le \imag\ \le 22.15$.
}
\label{fig_ms1054morfs}
\end{figure*}

We visually classified the morphologies of all galaxies in each field with $\imag\ \le 23.5$ 
(for the $z < 1$ clusters) or $\zmag\ \le 24$ (for the $z > 1$ clusters) regardless of 
their position or color. For reference, the characteristic magnitude, m$^*$, for cluster 
galaxies is $\imag\ = 22.3$ at $z = 0.83$ (Goto \etal 2004) and $\zmag\ = 22.7$ at $z = 1.24$ 
(Blakeslee \etal 2003b).  For all our cluster observations, we have at least one filter that 
samples part of the rest-frame $B-$band (see Table~\ref{tab_obssum}) so that morphological 
classifications can be readily compared with those at lower redshifts. 
We classify galaxies using the common Hubble sequence: E, E/S0, S0, S0/Sa,
Sa, Sb, Sb/Sc, Sc/Sd, Irr. However, for the purposes of the present analyses,
we bin these finer classifications into just 3 broad morphological categories: 
E (elliptical; $-5 \le T \le -3$), 
S0 (lenticular; $-2 \le T \le 0$),
and Sp$+$Irr (Spiral $+$ Irregular; $1 \le T \le 10$). The FWHM of the point spread 
function in our co-added ACS images is $\sim 0.09$ arcsec (1.8 WFC pixels), 
corresponding to a projected proper distance of 684 pc at $z = 0.831$ and 752 
pc at $z = 1.27$. We are thus able to resolve sub-kpc structure in all cluster 
members.  At \imag\ = 23.5, the typical galaxy subtends an isophotal area of 
$\sim$400 WFC pixels or 125$\times$ (FWHM)$^2$, making visual classification 
(or for that matter any classification method) quite feasible. At fainter 
magnitudes, however, classification rapidly becomes difficult and systematic 
errors increase both because galaxies are becoming smaller (e.g., Roche \etal 1998;
Bouwens \etal 1998; Ferguson \etal 2004; Trujillo \etal 2004) and because there is insufficient area over 
which the integrated S/N is sufficient for accurate classification. 
Examples of the ACS image quality and our corresponding classifications
are shown in Figures~\ref{fig_cl1604morfs} and \ref{fig_ms1054morfs}. 

\begin{figure*}
\plotone{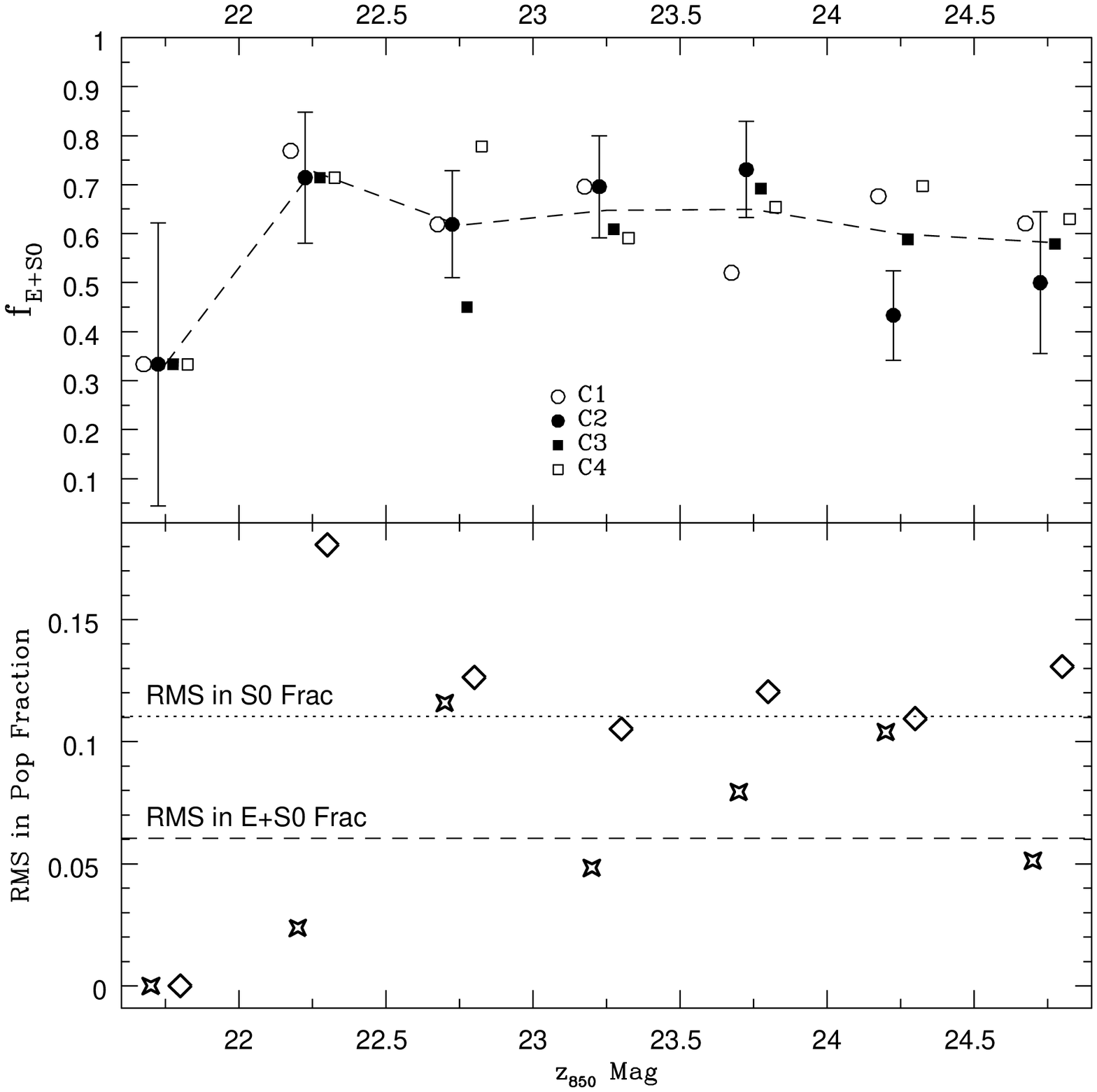}
\caption[]{
{\bf Upper panel}: The E+S0 population fraction as a function of \zmag\ magnitude for each of
the 4 independent visual classifiers. The sample used in this comparison is a color-selected sample
that favors inclusion of bulge-dominated galaxies, hence the relatively high overall
early-type population fraction. The values for each classifier are offset slightly from
one another for clarity. The error bars shown represent the $\sqrt{N}$ uncertainties.
The dashed line shows the mean E+S0 population fraction averaged over all classifiers.
{\bf Lower panel}: The RMS scatter in the E+S0 (stars) and S0 (diamonds) population 
fractions as a function of \zmag\ magnitude
between the 4 classifiers. The dashed lines represent the average RMS scatter.}
\label{fig_fracvseye}
\end{figure*}

The morphological classification was performed on the full sample of 4750 galaxies 
(in 7 clusters) by one of us (MP). Three other team members (NC, MF, BH) classified a 
subset of $\sim400$ of these galaxies to provide an estimate of the uncertainty in the 
classifications.  All classifiers used a common reference set of morphologies 
from a low redshift $B-$band galaxy sample as a guide. Exact or majority 
agreement between all 4 classifiers in the overlap sample was typically 
achieved for 75\% of the objects brighter than $\imag\ = 23.5$. Furthermore, 
there was no significant systematic offset between the mean classification for 
the 3 independent classifiers (as determined using the voting scheme from 
Fabricant \etal 2000) and the classification by MP giving confidence that the 
full sample was classified in a consistent manner. 

The overall population fractions between the 4 independent classifiers exhibit
only a relatively small variance. 
Figure~\ref{fig_fracvseye} shows the E+S0 fraction for each classifier 
as a function of \zmag\ magnitude for a color-selected ((\imag\ - \zmag) $\ge 0.5$)
subset of galaxies in the RDCS1252-2927 field. The average rms scatter in the E+S0 population
fraction between classifiers is $0.06$. The average rms scatter in the S0 population
fraction between classifiers is about 2 times higher, $0.11$. The rms in the E population
fraction is the same as that for the S0's, $0.11$.
In other words, the population fractions are fairly
robust and the variance in these fractions is significantly less than the 
$\sim 20 - 25$\% disagreement level between classifiers on the morphological classification of
any individual galaxy. Not surprisingly, the combined E+S0 fraction is more robust than
either the E or the S0 fractions -- a quantitative demonstration that detecting spiral structure
is a more robust skill than detecting disks. Most importantly, there are no significant systematic differences
between the classifiers. The scatter in the S0 population fraction is comparable with
$\sqrt{N}$ uncertainties in any given value and while counting statistics do not suggest the minimum scatter one
might expect between different classifiers, this level of scatter indicates that our classification errors are
small enough for the task at hand - providing a uniform set of classifications.

\begin{figure*}
\plotone{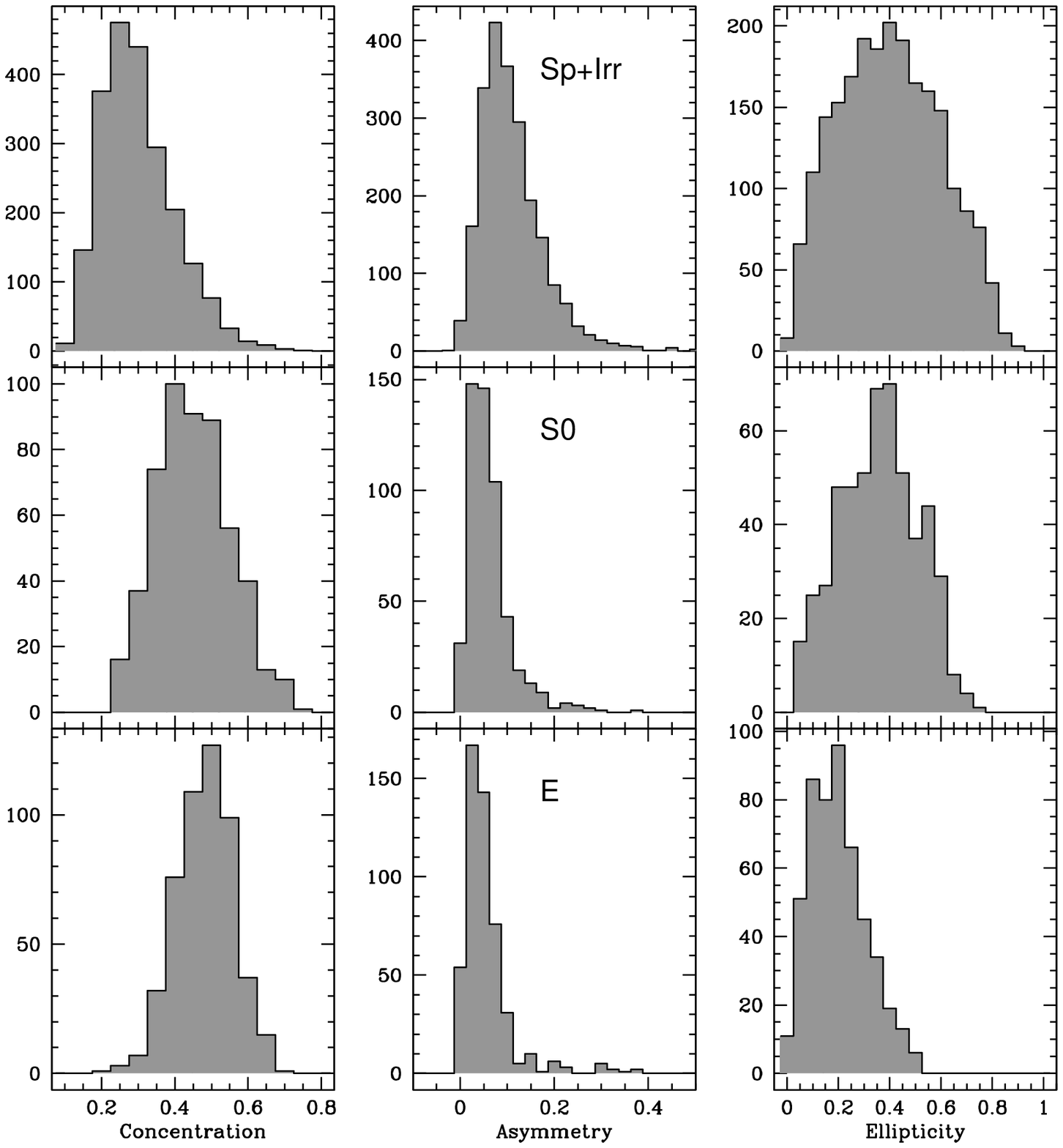}
\caption[]{
The image concentration, rotational asymmetry, and ellipticity distributions for galaxies
visually classified as elliptical, S0, and Sp+Irr. The top row shows the distributions for Sp+Irr.
The central row shows the S0 distributions. The bottom row shows the distributions for
ellipticals.}
\label{fig_acemorf}
\end{figure*}  

If our visual morphological classifications are robust, there should be noticeable differences
in the distributions of the objectively derived ``form" parameters (ellipticity, asymmetry, concentration)
for the E, S0, and Sp+Irr categories. Figure~\ref{fig_acemorf} shows the histograms of these
3 form parameters for each of the 3 morphological bins. Clear differences between the distributions exist. For
example, the ellipticity distribution of visually classified elliptical galaxies differs from that for visually
classified S0 galaxies at greater than the 99.999\% confidence level. 
We will use this fact later on as a key part of our analysis.
Ellipticals as a class have, as expected, a lower
median ellipticity and asymmetry and a higher median concentration than the other two morphological classes.
Ellipticals also exhibit less scatter about the mean values of these form parameters. The Sp+Irr class exhibits a
higher mean ellipticity and asymmetry than either the E or S0 class. The S0 galaxies have form characteristics
that are, on average, intermediate between the E and Sp+Irr distributions.

Two key concerns when performing galaxy morphological classification, visually or via 
machine algorithms, are the effect of surface brightness dimming and the shorter rest-frame 
wavelength being imaged with increasing redshift. The latter effect, sometimes referred to as 
the ``morphological k-correction," has been studied fairly well 
(Bunker \etal 2000, Abraham \& van den Bergh 2001, Windhorst \etal 2002, Papovich \etal 2003). 
In general, over the redshift range being
studied here, the morphological k-correction has been shown to be important only in a small 
fraction ($\ls 20$\%) of the galaxy population and in those cases, it is often more 
of an issue of how one characterizes any existing spiral structure and not usually a case of 
missing spiral structure altogether (see references above for details). Furthermore, the bluest wavelengths
we use for our morphological classifications correspond to the blue end of the rest-frame $B-$band,
which is where many lower redshift studies have been done. Our classifications are never
performed in the rest-frame $U-$band.

\begin{figure*}
\plotone{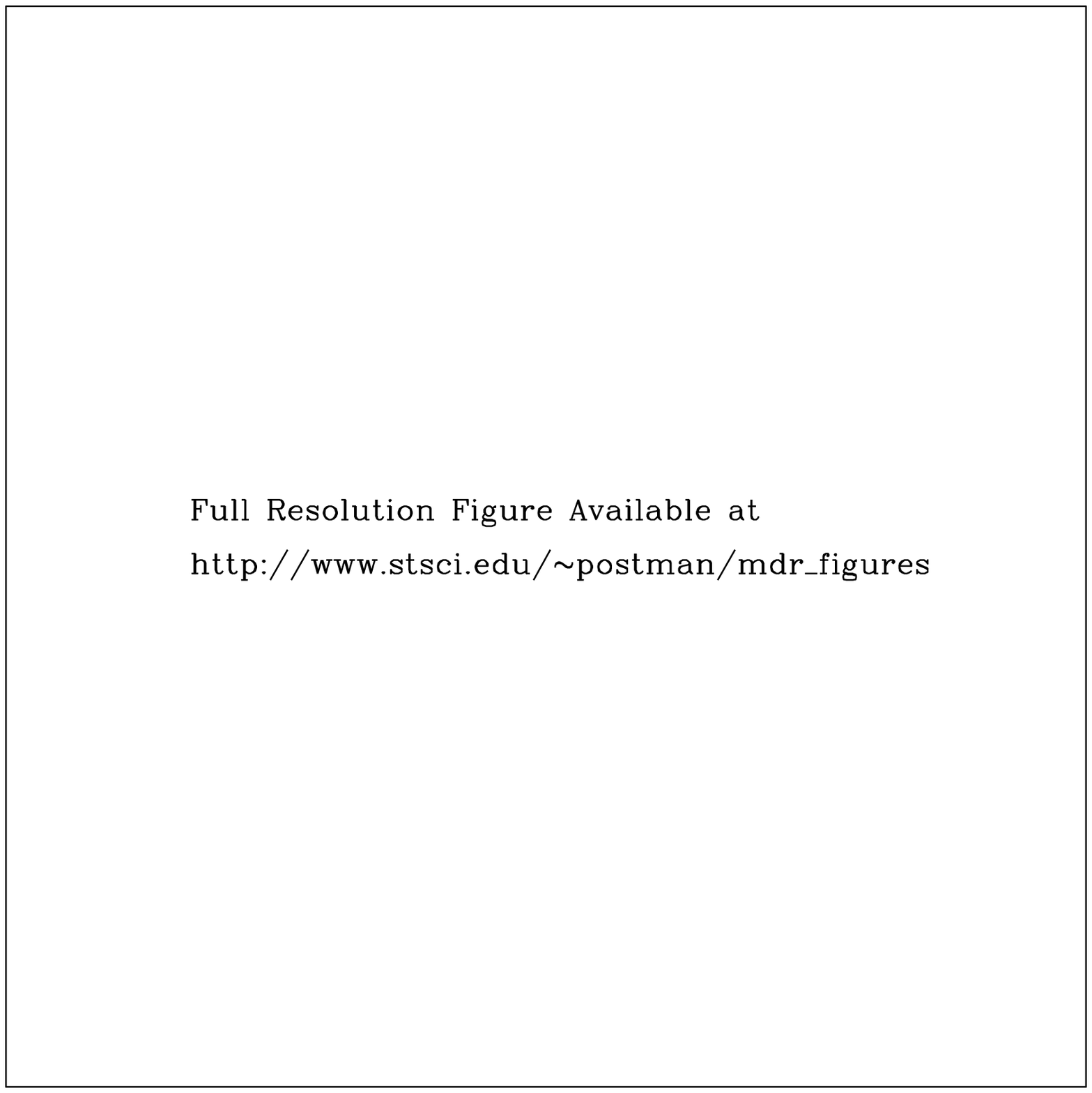}
\caption[]{
Image cutouts of 21 original ($z=0.83$) and redshifted ($z=1.24$) MS1054-0321 
galaxies and their corresponding morphological classifications. The original image is on the left. 
The galaxies shown here are spectroscopically confirmed cluster members. 
The redshifted images are constructed to match the exposure level used for 
our \zmag\ mosaic of RDCS1252-2927 (see Table~\ref{tab_obssum}).}
\label{fig_z083vsz124}
\end{figure*}

\begin{figure*}
\plotone{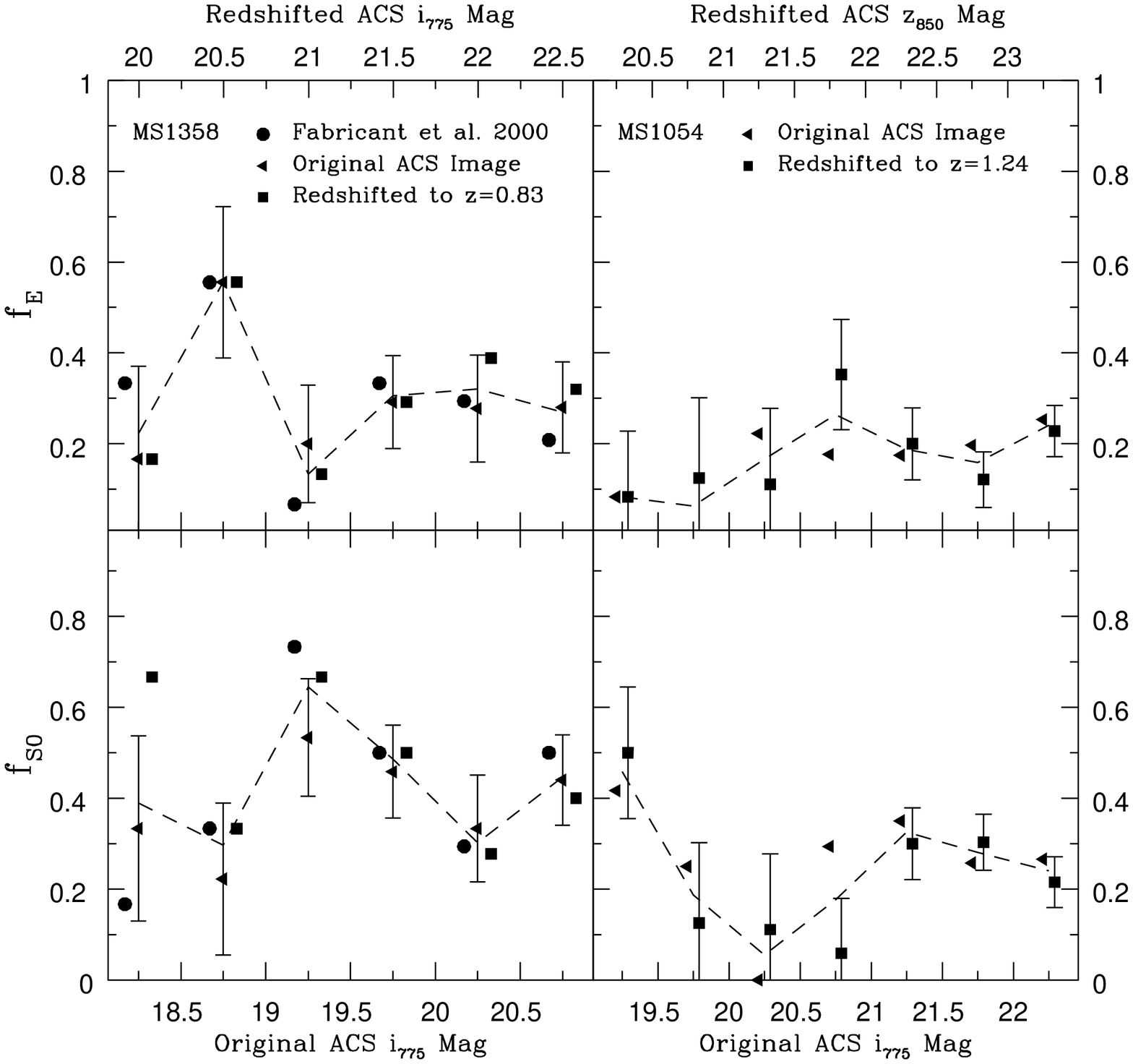}
\caption[]{
{\bf Left Panels:} The population fractions of E and S0 galaxies as a function of \imag\ magnitude
for the cluster MS1358$+$6245 at $z=0.33$. The population fractions shown are from 
Fabricant \etal (2003) (based on their WFPC2 mosaic), and from this paper (based on 
the \imag\ ACS WFC image and a redshifted version of this image out to $z=0.83$). 
There are no obvious systematic offsets between
our fractions and those in Fabricant \etal nor any systematic changes as we redshift the data out
to $z=0.83$. The dashed lines show the population fractions obtained by averaging the results
of all classifiers.
{\bf Right Panels}: The population fractions in the original MS1054-0321 ACS image 
and in a version redshifted to $z=1.24$. The population fractions at each magnitude are 
offset from one another by a small amount for clarity.}
\label{fig_zpopfrac}
\end{figure*}

The effects of SB dimming are potentially of greater concern as the ability to distinguish 
between adjacent categories (\eg E vs S0, S0 vs Sa) may be compromised at higher redshift. 
To test how sensitive our visual classification scheme
is to SB-dimming, we performed two simulations. In the first test, we ``redshift" our ACS image 
of the $z=0.33$ cluster MS1358$+$6245 to $z=0.83$ and perform visual classifications on 
both the original and redshifted versions. In the second
test, we redshifted our ACS image of MS1054-0321 to $z=1.24$ and compared the 
classifications derived for the original and redshifted versions. The ``redshifting" 
involved dimming the objects appropriately, resampling the
images to account for the smaller angular scale, and adjusting the noise levels to correspond 
to those appropriate for our exposure times used in the more distant cluster observations. 
Reclassification of the redshifted galaxy images was done in a random order and at least 
3 months after the initial classifications to minimize ``memory" of the initial classifications 
by the classifier. Some examples of the original and redshifted MS1054-0321 galaxies
are shown in Figure~\ref{fig_z083vsz124}. The results of the comparisons between the morphological 
classifications of the original and redshifted objects are shown in Figure~\ref{fig_zpopfrac}.
The population fractions obtained using the redshifted images are completely consistent with
those in the original images -- the differences are comparable to or less than 
the $\sqrt{N}$ uncertainties. Thus, our increased exposure times for the more
distant clusters coupled with the high angular sampling and sensitivity of the ACS/WFC 
successfully mitigate the effects of SB-dimming and allow us to distinguish between
our 3 primary morphological categories (E, S0, and Sp+Irr) uniformly across the redshift range
under study.

\subsection{Classification of Mergers \& External Morphology Comparisons}\label{mergerclass}

\begin{figure*}
\plotone{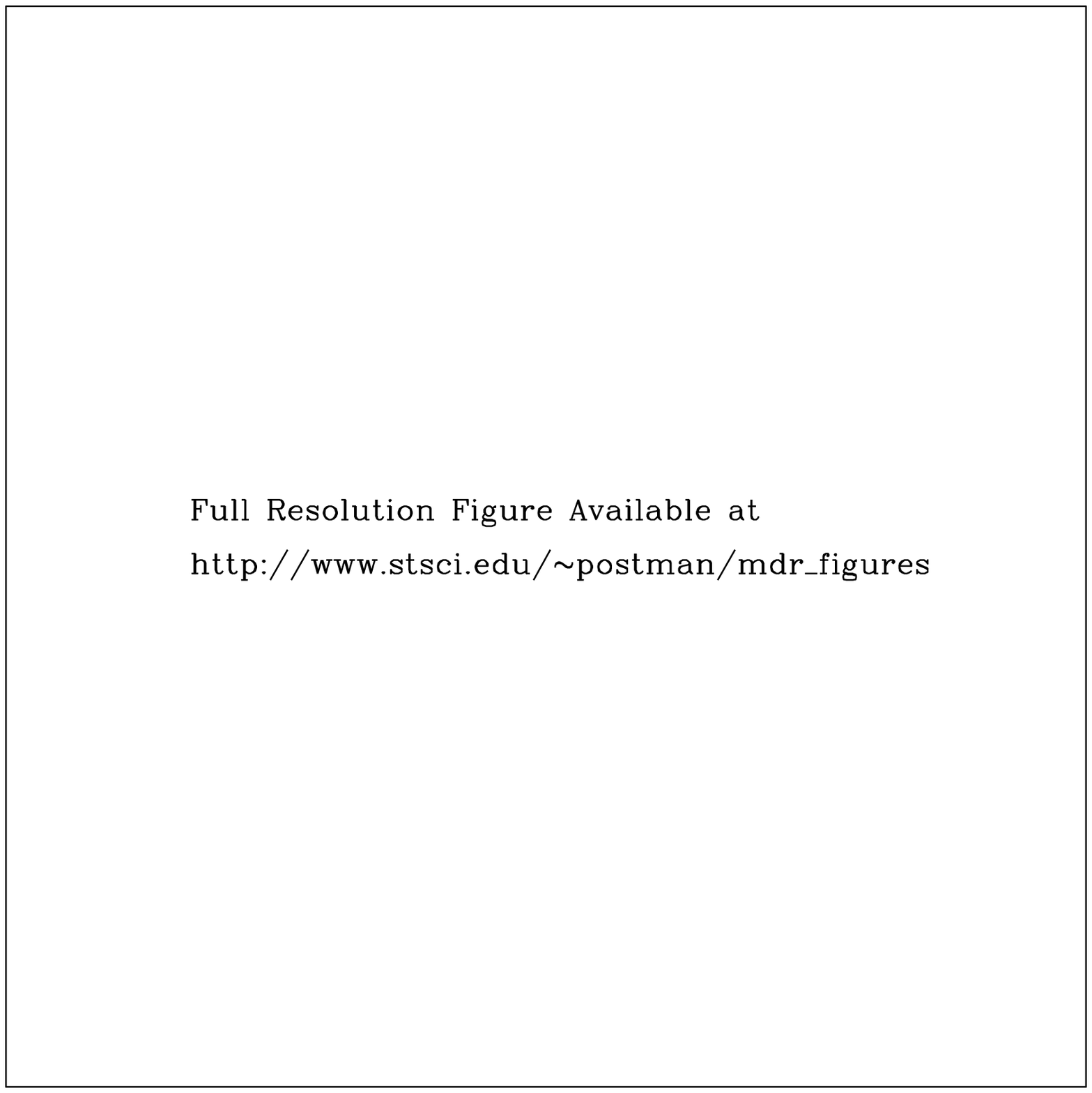}
\caption[]{
Color composite images of the 16 galaxies in common between our morphological sample
for MS1054-0321 and those classified as Merger/Peculiar by 
van Dokkum \etal (2000).  Our morphological classification of the object centered in each
cutout is indicated in the lower
left corner. The pictures here are
made from the \Vmag, \imag, and \zmag\ ACS/WFC images.}
\label{fig_vdmergers}
\end{figure*}

As we wish to measure the MDR and MRR in forms that can be compared to previous work,
we attempt to provide a standard Hubble type classification (E, S0, or Sp+Ir) for all galaxies
above our flux limits.
We do make a separate note on whether the object appears to be undergoing a merger or tidal
disruption and the analyses of the distribution and frequency of such systems will be
presented in a separate paper (Bartko \etal 2005). However, for the present work,
we do not classify galaxies as merger/peculiar systems (as
done by van Dokkum \etal 2000; hereafter referred to as vD2000) if one of the above standard morphological 
categories can indeed be applied to the individual objects involved in the merger. 
Nonetheless, the WFPC2 study of MS1054-0321 by 
vD2000 provides an additional check on the robustness of our
morphological classifications.  There are a total of 79 galaxies in common that have
morphological classifications by us and by vD2000. 
Of these, 16 are classified as $M/P$ (merger/peculiar) by vD2000. 
We classify 13 of them as bulge-dominated systems (E, S0, or S0/Sa) and 3 as later-type spiral galaxies. 
The ACS cutouts of these 16 objects are displayed in Figure~\ref{fig_vdmergers}.  
As can be seen from this figure, the morphology of most of the 
systems classified as mergers/peculiar by  vD2000 
can also reasonably be placed into one of the E/S0/Sp+Irr bins. 
While this confirms the vD2000 conclusion that MS1054-0321 
hosts a significant fraction of early-type mergers, it
does yield one difference (albeit perhaps a semantic one) in the conclusions reached regarding the overall
early-type populations in MS1054-0321. By counting mergers/peculiar systems as a separate
category, vD2000 concluded that early-type systems comprise a
lower population fraction (44\%) than in comparably rich clusters at lower redshift. 
We conclude that the early-type fraction, $\feso$, in MS1054-0321 is higher -- about 73\%, 
when one attempts to classify cluster members (including merger components) as E, S0, or Sp+Irr. 
Of course, the factor is a function of local density and 
the above value is averaged over densities in the range $15 < \Sigma \le 1000$
galaxies Mpc$^{-2}$. 
Given that the merger/peculiar
category has not routinely been used in the classification of low-$z$ clusters, we feel
it is an important exercise to provide a set of morphological classifications that
are as similar as possible to the low-$z$ studies. We also agree, however, that
quantifying the frequency of mergers at higher redshift reveals 
fundamental information about the evolution of cluster galaxies. 
The  assessment of the early-type population component of MS1054-0321 by us and vD2000 
are ultimately consistent, however, if one accounts for the observation that the majority of the close 
pairs in MS1054-0321 include at least one bulge-dominated member.

We compare our Hubble classifications with those galaxies that vD2000 
did classify as E, S0, or Spiral/Irr (i.e., excluding the M/P objects). 
We find exact agreement with their Hubble 
classification (when binned into these 3 categories) 71\% of the time. We swapped E or 
S0 classifications 11\% of the time (i.e., they called it E and we called it S0 or vice versa) 
and we swapped S0 and Spiral classifications 10\% of the time. The remaining 8\% were  
cases where either we or they could not make a definitive classification.
This translates to a $\pm0.1$ scatter between our respective $\fe$ or $\fso$ values, which
is consistent with the scatter estimated from our comparisons between our ACS
team classifiers.

As a further external validation of our morphological classifications, 
MP classified all galaxies from our ACS exposure of MS1358$+$6245 
($z=0.33$) that were in common with the extensive study of this system 
performed by Fabricant \etal (2000). Agreement between the MP classifications 
and those from Fabricant \etal was achieved $\sim 80$\% of the time
with no systematic bias seen in the discrepant classifications (see Figure~\ref{fig_zpopfrac}).
We thus conclude that our E, S0, and Sp+Irr classification scheme is robust and produces
Hubble types that are comparable in accuracy to visual morphological data used in other studies.

\subsection{Field Morphological Population Fractions}

\begin{figure*}
\plotone{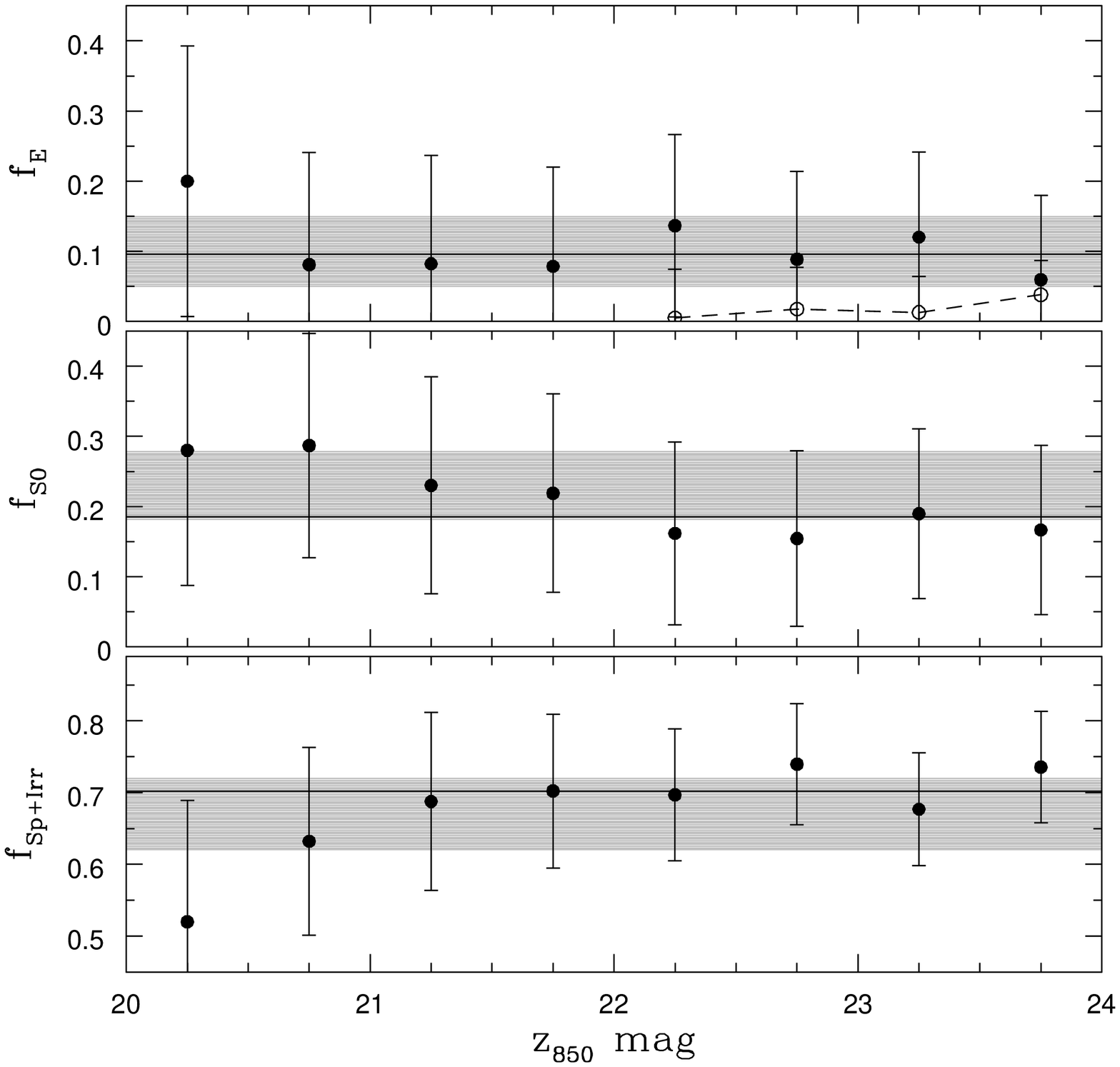}
\caption[]{
The population fractions of E, S0, and Sp+Irr of field galaxies (in our survey) as a function of \zmag\ magnitude.
The horizontal lines show the mean values over the range $20 \le \zmag\ \le 24$. The grey
shaded regions denote the local ($z \ls 0.15$) population fractions at low density ($\Sigma \ls 0.1$
galaxies Mpc$^{-2}$). The open circles in the upper plot show the fraction of galaxies
for which classifications could not be readily made. 
}
\label{fig_morfvsmag}
\end{figure*}

Figure~\ref{fig_morfvsmag} shows the E, S0, and Sp+Irr fractions
as a function of \zmag\ for galaxies either in the field. The data points are the
fractions derived from our ACS data and our visual classifications.
The results in this figure are based exclusively on galaxies with spectroscopic
or photometric redshifts that are incompatible with their being cluster members. 
The grey bands show the typical range in low-density population fractions derived from local
galaxy redshift surveys (\eg PG84, D97, Goto \etal 2003a). 
The local ($z < 0.1$) and distant ($0.5 \simless z \simless 1$) field (\ie low-density) galaxy populations appear to have 
similar fractions of E, S0, and Sp+Irr systems.

\section{Local Projected Galaxy Density Estimation}\label{density}

We compute the local projected galaxy density in two different ways to ensure robustness - the
nearest N neighbors approach used by D80 and D97 and a friends-of-friends algorithm.
Both methods yield consistent results and we therefore present our results in
terms of the nearest N neighbors based density unless otherwise specified.
Appendix~\ref{app_compden} provides a demonstration of the consistency of these
two density estimation techniques.
In the nearest neighbor method, one computes the area of the region containing
the N nearest neighbors and then derives the corresponding projected density
at each galaxy location from the expression:
\begin{equation}
\Sigma_i  = {f_{corr}(M_{CL},M_{Ref}) \over \left( \Omega_{N} D_{A}^2 \right)} 
\left( \sum^{N+1}_{k=1}(w(m_k,c_k)^{-1}) - N_{bkgd}\right) \label{eq_density}
\end{equation}
where $\Sigma_i$ is the projected galaxy density about a given
galaxy, $f_{corr}(M_{CL},M_{Ref})$ is a correction
factor that ensures the density is always measured with respect to a common fiducial
luminosity that corresponds to that used in low-redshift studies of the MDR, 
$N$ is the number of nearest neighbors, $w(m_k,c_k)$ is the
selection function (that can depend on magnitude and color -- see Appendix~\ref{app_popfrac}
for details), $ N_{bkgd} $ is 
a background contamination correction (if needed), $\Omega_{N}$ is the solid angle
of the region containing the N nearest neighbors (a rectangular region in our
implementation), and $D_{A}$ is proportional to the angular diameter distance
to the cluster (essentially the conversion between arcsec and projected Mpc).
The correction factor is
\begin{equation}
f_{corr}(M_{CL},M_{Ref}) = {\int_{-\infty}^{M_{Ref}} \Phi(M) dM \over
                            \int_{-\infty}^{M_{CL}} \Phi(M) dM} \label{eq_fcorr}
\end{equation}
where $M_{Ref}$ is the absolute magnitude limit to which we measure all
densities, $M_{CL}$ is the available limit for the cluster being analyzed,
and $\Phi(M)$ is the galaxy luminosity function (with $M^{*}_{V}(z=0) = -21.28$
and $\alpha = -1.22$).
We choose $M_{Ref}$ to match that of the original D80 study 
at $z = 0$ ($M_{V} = -19.27$ for our
adopted cosmological parameters) but as we
are sampling lookback times over which significant evolution in the
characteristic magnitude of a Schechter luminosity function is detected,
we allow $M_{Ref}$ to vary with redshift as $M_{Ref}(z) = M_{Ref}(z=0) - 0.8z$.
For our data, $f_{corr}$ lies in the range [1.2,3.0]. We use the density derived
from the $N=7$ nearest neighbors but the results are not sensitive to
this choice in the range $5 \le N \le 10$.

The background correction is applied only when we are using samples requiring a 
statistical background subtraction. For samples based on spectroscopic or photometric
redshifts, the background subtraction (if needed at all) is computed using the prescriptions 
described in Appendix~\ref{app_popfrac}. Our statistical background correction, $N_{bkgd}$, 
is derived from a combination of ACS and ground-based data. We use ACS observations of the HDF and 
Tadpole Galaxy (Ben\'{i}tez \etal 2004) to generate the surface density of field galaxies 
when $\imag\ > 23.0$.  For $\imag\ \le 23.0$, we transform the number counts from the large 
$I-$band survey of Postman \etal (1998b) to the required ACS bandpasses. 

Density estimation is most accurate when using galaxy samples that are 
based on spectroscopic or photometric redshift information as fore/background objects
are effectively excluded from the analyses. In cases where sufficient redshift information
is not available, one can use statistically--subtracted background corrected density estimates.
Such estimates are only reliable in dense
regions ($>80$ galaxies Mpc$^{-2}$) where the cluster population dominates the counts. 
For reference, the $1\sigma$ error in the field galaxy surface density at \zmag=24 is
5.4 galaxies arcmin$^{-2}$, which
corresponds to a projected density of 26 and 22 galaxies Mpc$^{-2}$ at 
$z = 0.83$ and $z = 1.27$, respectively. The statistically-subtracted background 
corrected population fractions and densities can be biased if there happens to
be a significant overdensity in the line of sight to the cluster. Thus,
the most reliable results are those based on samples with complete or nearly complete 
spectroscopic or photometric redshift information. 
In Appendix~\ref{app_compden} we demonstrate that a reliable
composite MDR or MRR can be derived from a combination of spectroscopic samples,
photo-z samples, and samples with statistically subtracted background corrections providing
each such sample is limited to its optimal density regime.

\section{The Morphology -- Density and Morphology -- Radius Relations at $z \sim 1$}\label{mdrresults}

\begin{figure*}[ht]
\plotone{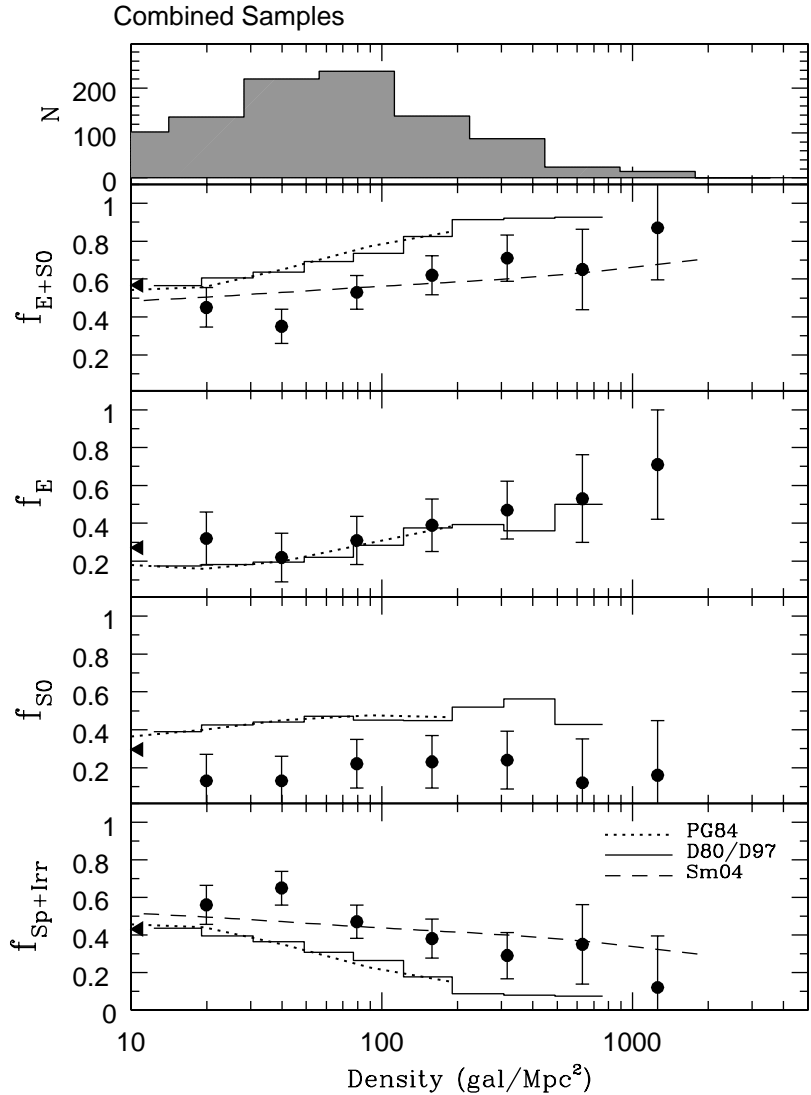}
\caption[]{
The MDR for all clusters based on the best available data. See text for details.
Errorbars include the uncertainties in counting statistics and morphological classification.
The previous results from low-$z$ surveys (D80, D97, PG84, SDSS)
and from $z \sim 1$ (Sm04) are shown for reference. The triangular markers along
the y-axis show the low-density population fractions from the SDSS (Goto \etal 2003a).}
\label{fig_mdr_best}
\end{figure*}

\begin{figure*}[ht]
\plotone{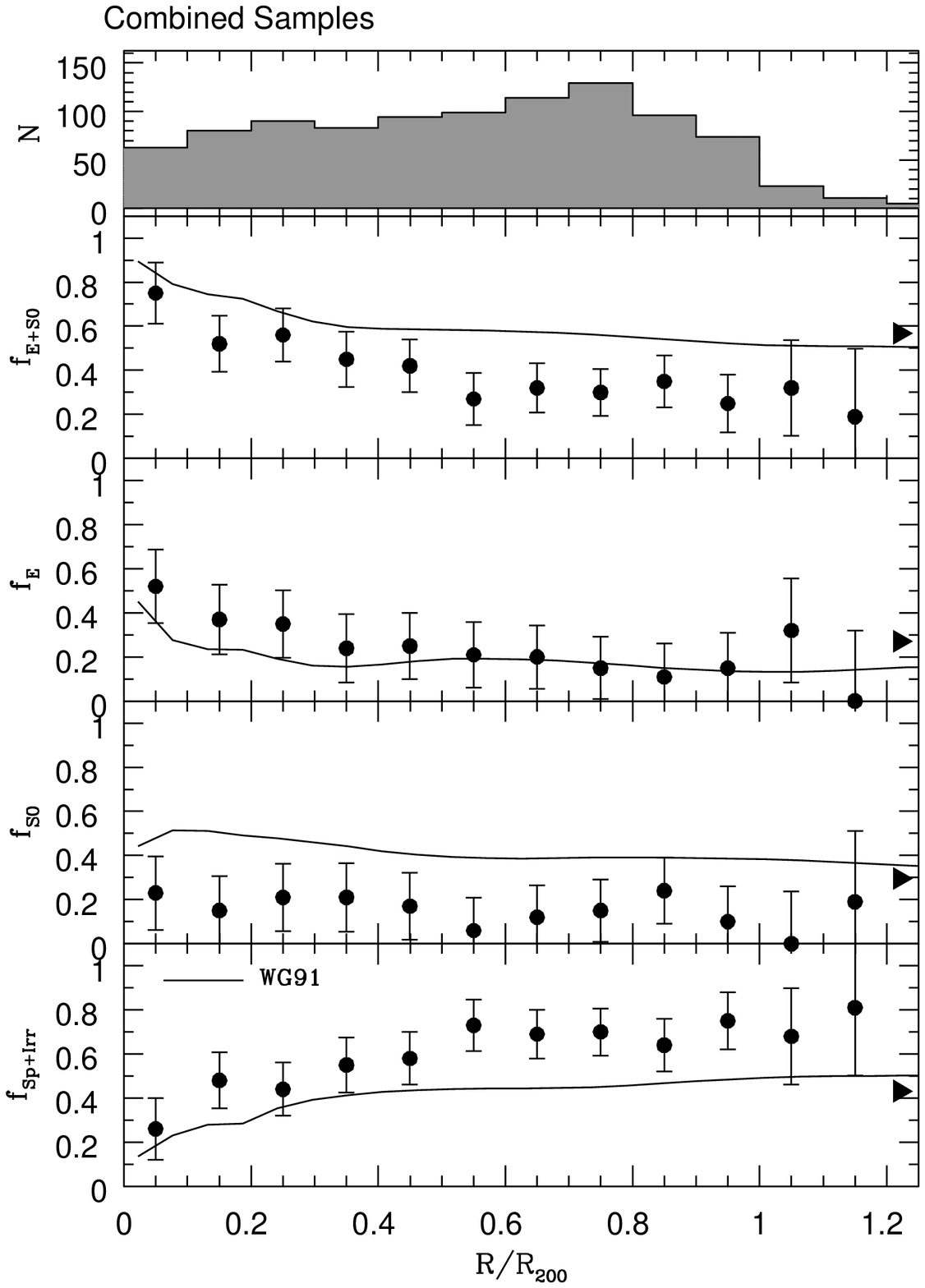}
\caption[]{
The Morphology -- Radius relation for all clusters based on the best available data. See text for details.
Errorbars include the uncertainties in counting statistics and morphological classification. Cluster
centers are determined from the centroid of the X-ray surface brightness distribution except for
CL1604+4321, where the centroid of the distribution of confirmed spectroscopic members is used instead.
The solid line is the low-$z$ morphology -- radius relation from Whitmore \& Gilmore (1991), converted to
\rtwoh\ units assuming a mean cluster redshift of $z = 0.04$ and a mean cluster velocity dispersion of 750 km s$^{-1}$.
The triangular markers along the y-axis show the low-density population fractions from the SDSS (Goto \etal 2003a).}
\label{fig_mrr_best}
\end{figure*}

We present our MDR  and MRR results in Figures~\ref{fig_mdr_best} through~\ref{fig_txvsfe}. 
Figure~\ref{fig_mdr_best} shows the composite MDR derived from all clusters in the sample using
the best available data. The composite MDR is derived from the spectroscopic
samples for MS1054-0321 and RXJ0152-1357 for densities below 1000 galaxies Mpc$^{-2}$ and
their photo-z selected samples for densities above 1000 galaxies Mpc$^{-2}$, 
the photo-z sample for RDCS1252-2927 (for all densities),
and the statistically subtracted background results for CL1604+4304,
CL1604+4321, RDCS0910+5422, and RXJ0849+4452. 
The statistically subtracted background results are used
only when $\Sigma \ge 80$ galaxies Mpc$^{-2}$ (roughly 3 times the
amplitude of the typical fluctuations in the surface density of fore/background
galaxies; see section~\ref{density}). As in Sm04, we find that the MDR exists at $z \sim 1$. 
The $z \sim 1$ MDR results from Sm04 are shown for comparison. Our $\feso$ vs. local density
results are consistent with those of Sm04 to within the $1\sigma$
uncertainties: we also find a 
less rapid increase in $\feso$ with increasing density than is seen at low redshift.
However, the elliptical fraction, $\fe$, shows no significant departures from the low
redshift $\fe$--density relation -- although, as demonstrated in Figure~\ref{fig_fracvseye}, 
the classification uncertainties in our $\fe$ and $\fso$ values 
are about twice as high as the uncertainties in our $\feso$ and $\fsp$ values.
The most notable difference between our current results and those at $z \simless 0.2$
(\eg see Figure~9 in Fasano \etal 2000) is the significantly lower fraction of S0 galaxies: averaged
over all densities with $\Sigma \ge 30$ galaxies Mpc$^{-2}$, our mean
$\fso = 0.20 \pm 0.12$ where the error includes both the uncertainties from counting
statistics ($\pm0.035$) and classification errors ($\pm0.11$). 
The typical low-$z$ $\fso$ value, when averaged over the same
density range, is $0.46 \pm 0.06$ (D80, D97, PG84) -- a factor of $\sim 2$ higher than what is found at $z \sim 1$.
A significant decline in the S0 population, relative to that seen in the current epoch, 
has previously been reported at redshifts as low as $z \sim 0.4$ (D97, Fasano \etal 2000).
Furthermore, over the range of densities being probed in this
study, $\fso$ exhibits only a weak dependence on the projected density, analogous to
what is seen at similar projected densities at lower redshifts (\eg D97).
The shallower growth of $\feso$ with increasing
density seen at $z \sim 1$ by Sm04 and by us thus appears to be due to a significant deficit of S0 galaxies
and an excess of spiral galaxies relative to similar environments in the current epoch.
This provides further support for observations suggesting that 
it is the S0 and spiral population fractions that are experiencing
the most significant changes with time over the past 8 Gyr (\eg
Moss \& Whittle 2000; Fasano \etal 2000; Kodama \& Smail 2001;
Treu \etal 2003; Sm04).

Figure~\ref{fig_mrr_best} shows the morphology -- radius relation (MRR) for the same composite cluster
sample used in deriving the MDR in Figure~\ref{fig_mdr_best}. For each cluster we have computed an
estimate of \rtwoh, the radius containing a mean overdensity of 200 relative to a critical universe, based on 
equation 8 in Carlberg, Yee, \& Ellingson (1997). The derived \rtwoh\ values are listed in Table~\ref{tab_xray}. For
RDCS0910+5422 and RXJ0848+4452, we assume $\sigma = 750$ km s$^{-1}$. We still need to define a cluster
center for each system and for that we use the centroid of the X-ray surface brightness
distribution. X-ray imaging in which the hot ICM is detected with adequate S/N levels is available for 
all clusters except CL1604+4321. For CL1604+4321, we use
the centroid of the distribution of spectroscopically confirmed members. In the case of RXJ0152-1357, the
X-ray distribution shows two well separated peaks (Della Ceca \etal 2000) and we thus subdivide the data for 
this cluster into two separate samples - a NE and SW component - and measure the radial distance of cluster galaxies 
in each subsample relative to the nearest X-ray peak. The low-$z$ reference for the MRR is taken from
Whitmore \& Gilmore (1991), with their radii converted to \rtwoh\ units using our cosmology and assuming
a mean cluster redshift of $z = 0.04$ and a mean cluster velocity dispersion of 750 km s$^{-1}$. To aid in the comparison between
the MDR and the MRR, we provide the approximate projected density -- radius relationship 
${\rm log}_{10}(\Sigma/635) \approx -1.63 (r /\rtwoh)$,
which is derived from our data. This approximate relation is not particularly accurate for radii less than 0.2\rtwoh\ and 
the cluster-to-cluster variation about this relation is substantial (factors of $2 - 4$ variation in the projected density at
a given \rtwoh-scaled radius), which is not surprising given the asymmetric galaxy distributions
in many of our $z > 0.8$ clusters (\eg Figure~\ref{fig_bpzmap}).
The key features of the MRR at $z \sim 1$ are that
(1) the bulk of the transition from a $\fsp$ consistent with that in the field environment to its
minimum value occurs within 0.6 \rtwoh\ ($0.6 \times \rtwoh$ corresponds to physical radii of 550 kpc -- 1.1 Mpc for these clusters), 
(2) the $z \sim 1$ $\feso$ value, at a given radius, is systematically less than
the low-$z$ $\feso$ for (r$/\rtwoh) \simless 1.0$, and (3) the $\fso$ -- radius relation shows the most
significant difference from the current epoch relationship. All of these characteristics are consistent with
those inferred from the $z \sim 1$ MDR. Given the significant asymmetry of the galaxy distributions in some
of these clusters, however, it is likely that the MRR is being diluted and is, thus, not as clean an indicator of 
morphological population gradients as the MDR for this particular cluster sample.

\subsection{Evolution of the Morphology--Density Relation}
\label{evolution}

\begin{figure*}[ht]
\plotone{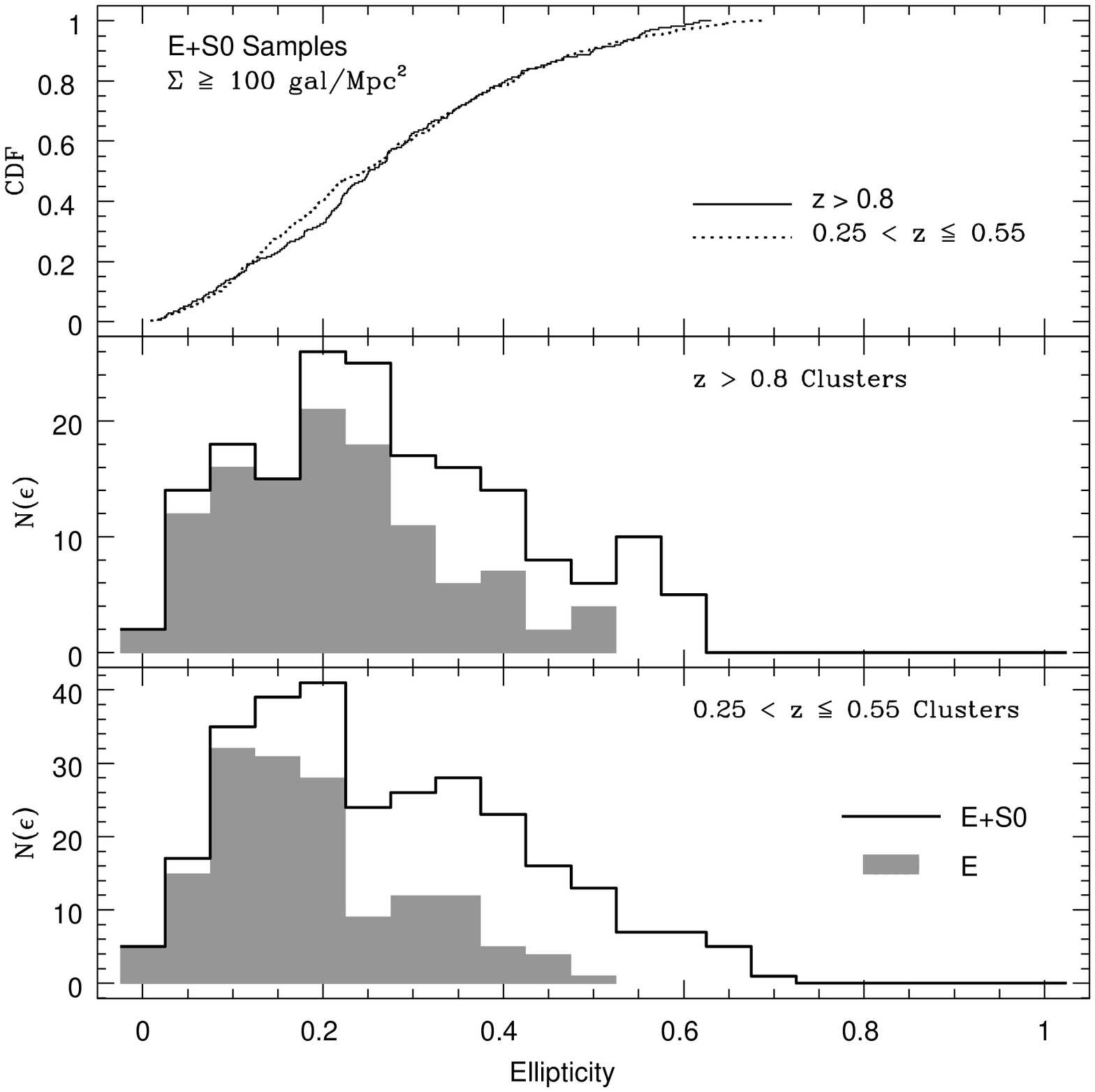}
\caption[]{
The distribution of ellipticities for the bulge-dominated galaxies that are likely cluster members
in 2 different redshift ranges. The lower redshift cluster data are also based on ACS imaging and come
from observations of 5 strongly lensing clusters in the range $0.25 \le z \le 0.55$. See section~\ref{discuss} for details.
The $z > 0.8$ galaxies are from our photometric redshift sample for the clusters
RXJ0152-1357, MS1054-0321, and RDCS1252-2927. The E+S0 ellipticity histograms are drawn with thick lines. 
The elliptical galaxy ellipticity distributions are represented by the
light grey shaded histograms. The uppermost panel shows the cumulative
distribution functions for the ellipticities of the E+S0 galaxies in each cluster sample.}
\label{fig_ellipvsmorf}
\end{figure*}

An additional way to assess the presence of S0 galaxies in clusters is to characterize the E+S0 ellipticity distribution,
as was done by D97. As shown earlier in Figure~\ref{fig_acemorf},
the ellipticity distributions of lenticular and elliptical galaxies differ substantially. For the full ACS galaxy sample shown
in  Figure~\ref{fig_acemorf}, the hypothesis that the E and S0 ellipticity distributions are drawn from the same
parent population is rejected at greater than the 99.999\% 
confidence level. Thus, if the $z \sim 1$ bulge-dominated
cluster galaxy population were truly devoid of a significant population of S0 galaxies, the ellipticity distribution of the
E+S0 galaxies would more closely resemble that of pure ellipticals and would not include a significant component of 
objects with ellipticities beyond 0.5.  Figure~\ref{fig_ellipvsmorf} shows the ellipticity distributions of the E+S0 
cluster galaxy populations for three of our $z > 0.8$ cluster galaxies along with galaxies from five clusters
with redshifts in the range $0.25 \le z \le 0.55$. 
To make a fair comparison of these two cluster samples, we must select objects that lie in similarly
dense environments. Therefore, the galaxies used in this comparison are all selected from 
environments where the local projected density is $\ge 100$ galaxies
Mpc$^{-2}$. The $z > 0.8$ sample used here consists of the photo-z selected cluster 
members in RXJ0152-1357, MS1054-0321, and RDCS1252-2927. We limit the $z > 0.8$ cluster sample to the three
clusters with good photo-$z$ data to ensure we are selecting probable cluster members.
The $z < 0.6$ cluster data
are from GTO ACS observations of 5 strongly lensing clusters in the range $0.25 \le z \le 0.55$ 
(HST PID \#9292). The 5 clusters are Zw1455$+$2232 ($z=0.258$), MS1008$-$1224 ($z=0.301$), 
MS1358+6245 ($z=0.328$), CL0016+1654 ($z=0.54$), and MS J0454-0300 ($z=0.55$).
Visual classification of all galaxies with $\imag \le 22.5$ in each $z < 0.6$ cluster ACS image was performed by MP. 
The rest-frame wavelengths
sampled by the \imag\ filter for $0.25 \le z \le 0.55$ typically lie in the $V-$band, thus any ``morphological" k-corrections 
between this sample and the $z > 0.8$ sample should be
small. A total of 798 galaxies were classified in the five $z < 0.6$ clusters. 
As the low-$z$ ACS images were all single pointings centered on the cluster core, the vast majority of the E and S0 galaxies 
identified are likely to be cluster members. For reference, the ACS/WFC subtends 800 kpc at $z = 0.26$ and 1.3 Mpc 
at $z = 0.55$.  A Kolmogorov-Smirnov test finds that the distribution of ellipticities of E+S0 galaxies in the
$z > 0.8$ clusters is inconsistent with being drawn from a pure elliptical ellipticity distribution at the 97.0\% 
confidence level. A Wilcoxon rank-sum test, which is better suited to 
measuring differences in the mean values of two distributions than a KS test, 
finds the E+S0 and E galaxy distributions for the $z > 0.8$ clusters
differ at the 3.1$\sigma$ level. We thus reject the hypothesis that there are no 
S0 galaxies in dense environments at $z \sim 1$. The E+S0 and E galaxy ellipticity 
distributions for the $0.25 \le z \le 0.55$ cluster sample are inconsistent with each other at the 99.998\% 
confidence level.

We can also apply a robust test ({\it i.e.}, one that primarily relies on our ability to distinguish only between
E+S0 and Sp+Irr) in an attempt to constrain the evolution of the S0 population
by comparing the $z > 0.8$ E+S0 ellipticity distribution with that for the $0.25 \le z \le 0.55$ cluster galaxy sample.
The cumulative distribution functions for the E+S0 ellipticities in the 
two different cluster samples are shown in the top panel in Figure~\ref{fig_ellipvsmorf}.
The ellipticity distributions for the $0.25 \le z \le 0.55$ and the $z > 0.8$ cluster samples are, however, completely 
consistent with one another (KS test rejects inconsistency only at the $\sim10$\% 
confidence level), suggesting that, at least in these particular samples, any evolution
in the relative E and S0 population fractions does not manifest itself as a significant difference in the E+S0
ellipticity distribution. It is possible that with larger and more homogeneously selected cluster samples over a range
of redshifts this test would provide a quite objective way to measure the evolution of the morphological population
fraction in dense environments.

\begin{figure*}[ht]
\plotone{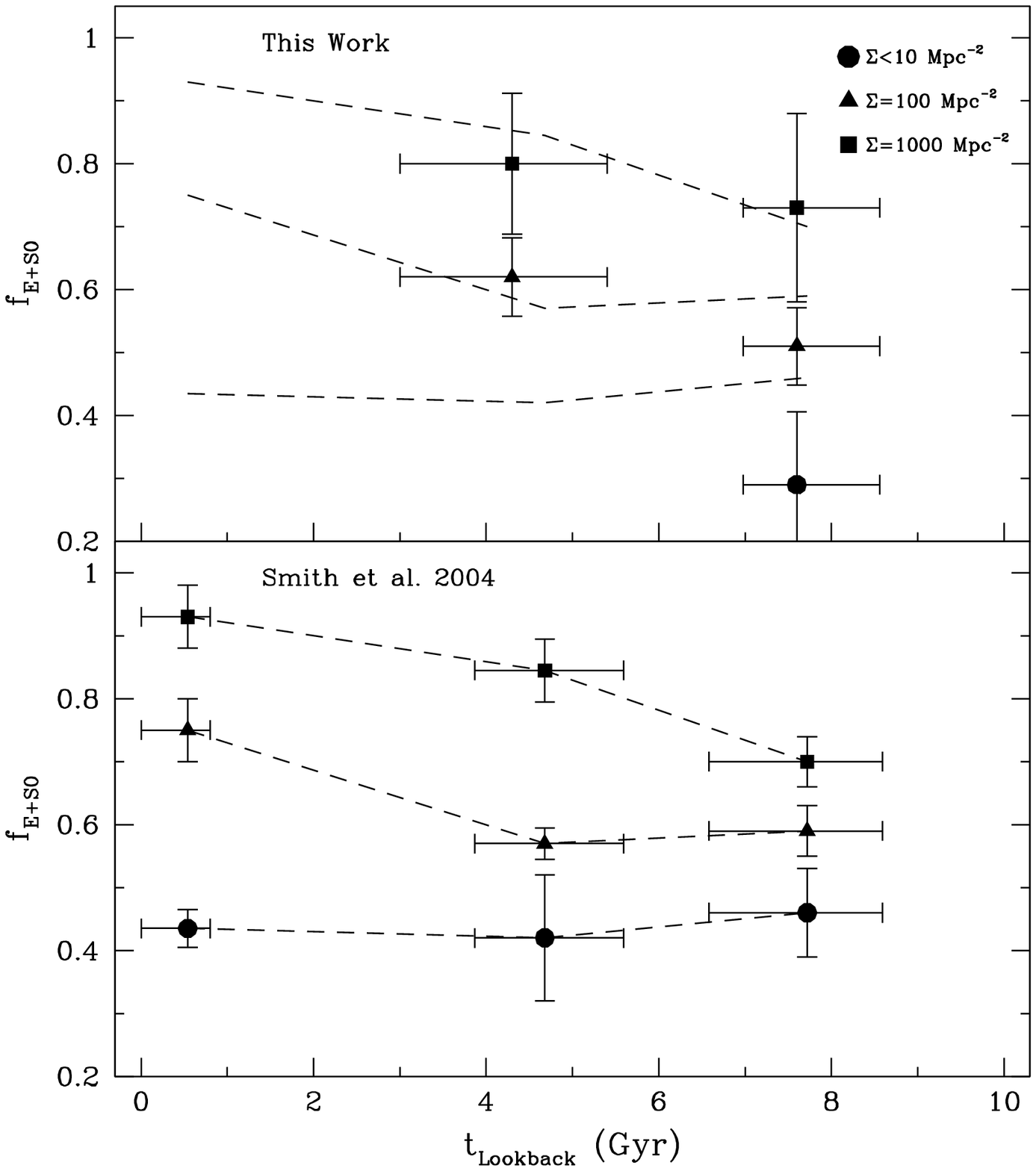}
\caption[]{
{\bf Bottom:} The evolution of the MDR reported by Sm04. Lookback times
have been transformed to compensate for the small difference between the
cosmology in Sm04 and that adopted in this paper - specifically,  
our choice of a slightly larger Hubble constant (70 vs. 65). The dashed lines representing
the change in $\feso$ with time at the 3 different density regimes are reproduced in all
three panels for reference. 
{\bf Top:} Our derived E+S0 fractions at projected densities corresponding to $<10$, 100,
and 1000$h_{65}^2$ galaxies Mpc$^{-2}$ (to match Sm04) for our combined $z \sim 1$ cluster sample and our
$0.25 \le z \le 0.55$ cluster sample. The latter sample is not suitable for measuring low-density population fractions.
}
\label{fig_fevst}
\end{figure*}

The $\feso$ as a function of lookback time and local density is shown in Figure~\ref{fig_fevst}. This figure
is modeled after Figure~3 in Sm04. The Sm04 results 
are reproduced for reference in the lower panel of Figure~\ref{fig_fevst}.
The upper panel of Figure~\ref{fig_fevst} shows the 
analogous results for our ACS $z > 0.8$ and $0.25 \le z \le 0.55$ cluster samples.
The results in the upper panel are based on the mean population fractions within logarithmic
density bins 0.4 units wide centered at 10, 100, and 1000$\ h^2_{65}$ 
galaxies Mpc$^{-2}$. 
Our $\feso$ values are in good agreement with the Sm04 values -  
any differences are comparable with the uncertainties.
We corroborate the key observational result of the Sm04 study that the most significant differences
between the MDR at low redshift and high redshift are confined to regions where the projected galaxy density
is larger than $\sim 40$ galaxies Mpc$^{-2}$. 
While the agreement is re-assuring given
the significant overlap of the clusters used in the two programs, 
and bolsters the concept of using visual morphological
classification in comparative studies at high-$z$, there is also significant room for reducing the
existing uncertainties that will be achieved only when a much ($\sim 10\times$) 
larger sample of $z \sim 1$ cluster galaxy morphologies is available.
Our derived population fractions as a functions of
projected density and radius are also provided in Tables~\ref{tab_mdr_values}~and~\ref{tab_mrr_values}
for easy reference.
As noted above, our results suggest that the observed MDR evolution
is primarily driven by evolution in the fractions of S0 and Sp+Irr galaxies.

\subsection{A Correlation between $\feso$ and $\lx$?}

\begin{figure*}[ht]
\plotone{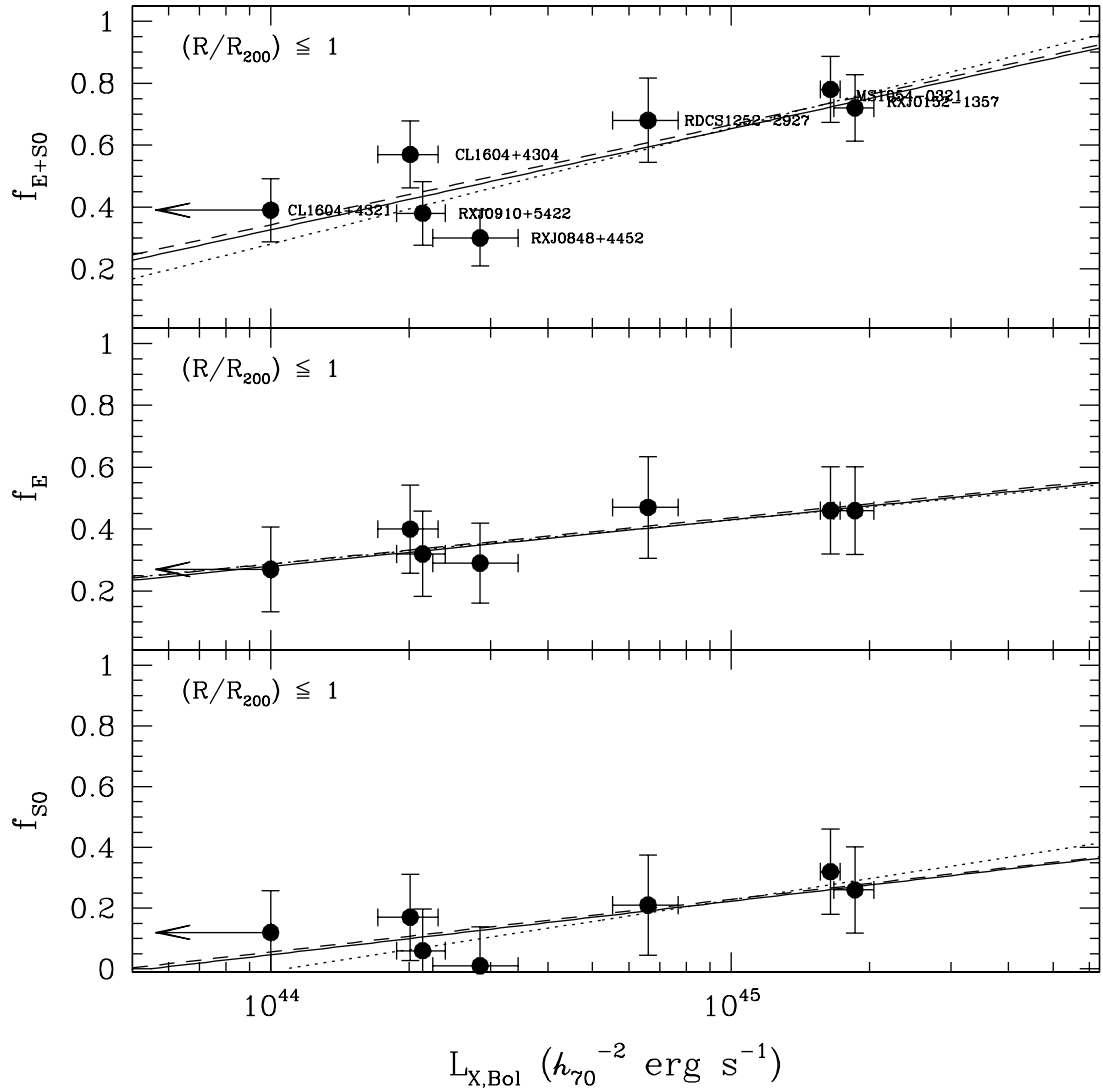}
\caption[]{
The dependence of the bulge-dominated galaxy population fractions, $\feso,\ \fe,\ {\rm and} \fso$, on the bolometric 
cluster X-ray luminosity within a radius corresponding to \rtwoh. 
Names of the individual clusters are shown in the upper panel.
The solid line is the best fit relation when the data are weighted
by the inverse square of their uncertainties. 
The dashed line is the best fit when each data point is given equal weight.
The dotted line is the best error-weighted fit with CL1604+4321 excluded. The data for
RXJ0152-1357 in this figure includes both the NE and SW components of the cluster. 
Errorbars include the uncertainties in counting statistics and morphological classification. }
\label{fig_lxvsfe}
\end{figure*}

Our sample exhibits a potentially interesting trend between the early type population fraction
and the cluster bolometric X-ray luminosity.
The bulge-dominated galaxy population fraction in clusters with high X-ray luminosity 
($\lxbol \ge 1.5 \times 10^{45}\ h^{-2}_{70}$ erg s$^{-1}$) is higher than that in
low X-ray luminosity ($\lxbol \le 3 \times 10^{44}\ h^{-2}_{70}$ erg s$^{-1}$)
clusters, even at the highest projected densities. This is demonstrated
in Figure~\ref{fig_lxvsfe}. Here we show $\feso$, $\fe$, and $\fso$ as a function of the bolometric
X-ray luminosity for the galaxies within \rtwoh\ for each cluster.
D80 found a modest increase in $\fso$, a corresponding
decrease in $\fsp$, and no change in $\fe$
in his sub-sample of 8 high X-ray emitting clusters relative to the full sample of 55 clusters. 
Least square, error-weighted, fits to the trends in Figure~\ref{fig_lxvsfe} give 
$$
\feso \propto \lxbol^{0.33\pm0.09}
$$
$$
\fe \propto \lxbol^{0.15\pm0.09}
$$
$$
\fso \propto \lxbol^{0.18\pm0.09}
$$
The linear correlation coefficients for the fits
to the log$(\lxbol) - \feso$, log$(\lxbol) - \fe$, and log$(\lxbol) - \fso$ data 
are 0.82 (97.6\% CL), 0.84 (98.2\% CL), 
and 0.75 (94.8\% CL), respectively. The correlations are significant at the $2 - 3\sigma$ level.

A correlation between $\feso$ and $\lx$ could, in principle, be produced either as a consequence
of environmental interactions subsequent to the formation of the cluster galaxies
and/or as a result of the initial conditions present at the time of their formation.
In a simplified example of the former scenario, the
ram-pressure, $P_{ram}$, acting on a galaxy moving through the intracluster
medium (ICM) is proportional to $\rho v^2$. If the ICM is in hydrostatic equilibrium,
then the bolometric X-ray luminosity, $\lx$,
is proportional to $\rho^2 \tempx^{1/2} R^3$ (\eg Ettori \etal 2004). 
The total mass of the system,
M$_{tot}$, is proportional to $\tempx^{3/2}$ and is also proportional to $v^2$.
The same family of scaling laws gives $\lx\propto \tempx^{2}$, although observationally
a steeper relation is found where $\lx\propto \tempx^{2.8}$ (\eg Ponman \etal 1996;
Mulchaey \& Zabludoff (1998); Xue and Wu (2000)).
These relations can be manipulated to yield $P_{ram} \propto \lx^{1.0 \pm 0.1}$, 
where the exponent value depends on the exponent in the $\lx - \tempx$ relation. 
The ram-pressure is stronger within more luminous X-ray clusters.
Therefore, if ram-pressure stripping were the dominant mechanism
responsible for the origin of S0 galaxies in clusters, $\feso$
should exhibit a positive correlation with $\lx$. However, many previous
observations, including the relatively weak dependence of $\fso$ on projected density, 
strongly suggest that ram pressure stripping alone cannot explain the 
morphological mix in clusters (\eg D80; D97; Kodama \& Smail 2001; 
Okamoto \& Nagashima 2003). 
 
Alternatively, a positive correlation between
$\feso$ and $\lx$ could be produced as a consequence of simple dynamics - 
the most massive (and, hence, most X-ray luminous) clusters will collapse earlier and any
environmentally-driven processes
that produce bulge-dominated galaxies will therefore have been active for a longer period
of time at any subsequent redshift. This simple timescale argument could result in a positive correlation
between the fraction of early-type galaxies and observational proxies for cluster mass,
regardless of whether or not the efficiencies of the transformation processes are correlated
with the properties of the ICM. However,
this would only be true if the timescales required to establish a significant population of early-type 
galaxies were long compared to the collapse time. If the timescale for establishing the 
early-type population were comparable to the collapse time (i.e., if deeper potential wells
are ``born" with a higher early-type population fraction), then the $\feso - \lx$
relation may be telling us more about cluster and galaxy formation processes than about the 
cluster evolution process.

\begin{figure*}[ht]
\plotone{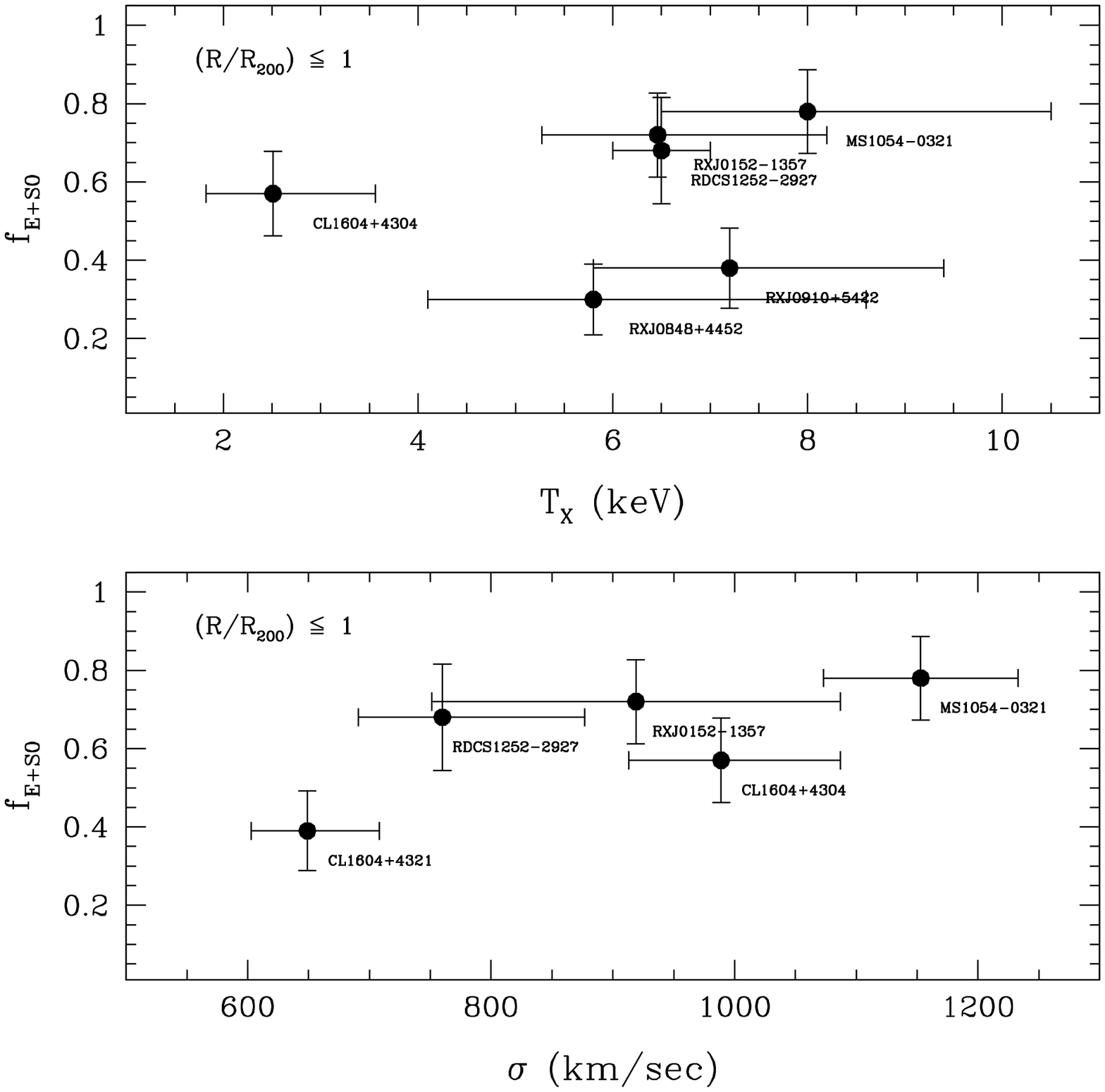}
\caption[]{
The bulge-dominated galaxy population fraction, $\feso$, within a radius corresponding to \rtwoh\  
as a function of the X-ray gas temperature and cluster velocity dispersion. 
Names of the individual clusters are shown.
Errorbars include the uncertainties in counting statistics
and morphological classification. }
\label{fig_txvsfe}
\end{figure*}

As the results in Figure~\ref{fig_lxvsfe} are based on only 7 clusters and as the significance of the
correlations are only significant at the $\simless 3\sigma$ level, we refrain from over-interpretation. 
Indeed, the correlations between $\feso$ and $\tempx$ and $\feso$ and cluster velocity dispersion, $\sigma$, are 
not significant (see Figure~\ref{fig_txvsfe}). 
In the literature, the mass dependence of 
cluster galaxy evolution has been a subject of debate.
For example, Fairley \etal (2002) studied eight X-ray selected clusters and did not find any
dependence of the blue galaxy fraction on $\lx$ -- a trend that might be expected if
$\feso$ correlated with $\lx$. De Propris \etal (2004), using
a larger sample of 60 clusters from the 2dF Galaxy Redshift Survey,
finds that the blue galaxy fraction does not depend on the 
velocity dispersion of the cluster galaxies. However, there is tentative evidence that cluster
integrated star formation rates (\eg Finn \etal 2004) correlate with $\tempx$ and $\lx$ 
(Homeier \etal 2005), which is presumably a dependence on cluster mass. Furthermore, 
Zabludoff \& Mulchaey (1998) found that groups of galaxies exhibit a strong correlation between
the bulge-dominated galaxy fraction, $\feso$, and the group velocity dispersion. And while the specific relation
they found cannot be extended to very massive clusters (their predicted $\feso$ reaches a value of unity
at a velocity dispersion of  $\sim700$ km s$^{-1}$), it does suggest that groups of galaxies
have a positive correlation between $\feso$ and $\lx$ 
through the observational relation $\lx \propto \sigma^{4.3}$ (Mulchaey \& Zabludoff 1998). 
Margoniner \etal (2001) and Goto \etal (2003b) find that blue fractions of cluster galaxies are lower for
richer clusters, a result that is consistent with a positive $\feso - \lx$ relationship.  

The significance of the $\feso - \lx$ relation clearly needs to be studied
using a larger sample so that sub-division
by mass (or a suitable observational proxy for mass)  can be conducted for a far greater number
of clusters. The HST/ACS snapshot program of 73 homogeneously selected X-ray clusters (ID\#10152, M. Donahue, P.I.) 
should provide the sample needed to assess this relation in the range $0.3 < z < 0.7$. 
If a significant $\feso - \lx$ relation is ultimately found, one implication is that
the MDR and its evolution may be dependent on the total cluster mass.
Certainly on galaxy-mass scales, there is evidence that the
star formation process occurs more rapidly in systems with higher stellar mass
(\eg Heavens \etal 2004). For the present sample, correlations between the fraction
of early type galaxies and the X-ray properties of the clusters provide, at best, a strong inspiration to study
the trends with samples explicitly geared to investigate this relationship. 

\section{Discussion}\label{discuss}

As originally discussed by Tully and Shaya (1984), 
the measurement of the evolution in the MDR is a significant step in
understanding the relative roles of environment and initial conditions
in establishing morphological population gradients. The 
less rapid growth in the bulge-dominated galaxy population with increasing density
(or decreasing radius) that is seen at $z \sim 1$, implies
that environment must play an important role in the establishment of the current
epoch MDR. Sm04 explore a range of morphological transformation
scenarios based on the expression:
\begin{equation}
\fso_{,z=1} = \feso_{,z=1} - \fe_{,z=0.5} {N_{z=0.5} \over N_{z=1}} + {\Delta N_E \over N_{z=1}}
\end{equation}
where $\fso_{,z=1}$ is the S0 population fraction in clusters at $z = 1$, 
$\feso_{,z=1}$ is the E+S0 fraction in clusters at $z = 1$, $\fe_{,z=0.5}$ is the
elliptical fraction in clusters at $z = 0.5$, $N_{z=0.5} / N_{z=1}$ is the ratio of the number
of galaxies in clusters at $z = 0.5$ to that at $z = 1$, and $\Delta N_E$ is the change in the
number of cluster ellipticals between $z = 1$ and $z = 0.5$. By combining varying
amounts of infall (including infall of early-type galaxies), mergers, and cannibalism, Sm04
find that for reasonable levels of each process the $\fso$ at $z = 1$ is typically less than 0.1.
The exception is for the case where there is no infall and 10\% of the spiral galaxies
merge in pair-wise manner to form ellipticals. In this case, the model predicts $\fso = 0.18$ at $z = 1$. 
The degree to which our observations conflict with the predictions for very low S0 population
fractions in $z = 1$ clusters is, of course, directly tied to the uncertainty in our $\fso$ measurements.
Our mean S0 population fraction is inconsistent with being zero at about the 90\% 
confidence level, which is not strongly in conflict with the Sm04 predictions for $\fso_{,z=1}$. 
However, given that the ellipticity distribution of the E+S0 galaxies in the $z \sim 1$ clusters is
inconsistent with being drawn from a population dominated largely by pure elliptical
morphologies and given that we do get a mean $\fso$ of $0.20\pm0.12$ for $\Sigma \ge 30$ galaxies Mpc$^{-2}$, 
we investigate minor variations to the Sm04 scenario that can 
increase in the predicted S0 population fraction at $z \sim 1$. 
An obvious choice is to explore reasonable modifications of the value of $\fe_{,z=0.5}$. 
Sm04 adopted $\fe_{,z=0.5} = 0.6$. However,
we find that for $\Sigma \ge 100$ galaxies Mpc$^{-2}$ a value in the range
$0.4 - 0.5$ appears to be closer to what the data suggest. Decreasing $\fe_{,z=0.5}$ to
0.5 increases the range of the Sm04 predictions 
to $0.09 \le \fso_{,z=1} \le 0.28$ for the models considered. This would make their predictions and our
mean $\fso$ values consistent within $1\sigma$. However, doing so does change the
conclusions of Sm04 a bit - namely that $\sim$50\% of the lenticulars in clusters could already be in place
at $z = 1$ and that the remaining half form largely between $z \sim 0.5$ and the present epoch.

Different morphological transformation mechanisms operate 
in different environments and, thus, by identifying the radius or density where the morphology 
of cluster galaxies start to change, we can hope to identify the underlying 
physical mechanisms. The breaks in the MDR, first characterized by PG84 and more 
recently by Goto \etal (2003), suggest different processes are probably 
responsible for the origin of elliptical and S0 galaxies. For example,
ram-pressure stripping (Gunn \& Gott 1972) is only efficient within the cluster core (typically $\lesssim 250$ kpc) where 
the ICM density is high enough to allow the dynamic pressure to overpower the gravitational restoring force
of interstellar gas in a galaxy's disk. 
Figure~\ref{fig_mrr_best} indicates that the fraction of elliptical galaxies starts to 
increase inwards of 0.6 \rtwoh, or in terms of local galaxy density, 
at around 70 galaxies Mpc$^{-2}$ (Figure~\ref{fig_mdr_best}). 
This environment matches that where the population 
fraction of Sp+Irr galaxies starts to decline. 
The scale of 0.6 \rtwoh\ for the onset of the transition between the low-density field population and that
in high density regions is in agreement with previous low redshift observations
 (\eg Kodama \etal 2001; Treu \etal 2003; Goto \etal 2003b). 
At distances beyond 250 kpc or so, ram pressure
becomes much less effective as a morphological transformation process 
and thus other environmental processes must play
a significant role in the establishment of the MDR and MRR (\eg Treu \etal 2003).
Galaxy-galaxy merging, for example, is 
most effective when relative velocity of galaxies are comparable to the rotation of 
galaxy itself ($\sim 200$ km s$^{-1}$) and is thus likely to be most effective in regions
far from the cluster core. Galaxies beyond 5 Mpc from the cluster center will, at $z \sim 1$, not yet have had
sufficient time to traverse the cluster core and should, therefore, not experience
tidal stripping or tidal triggering of the star formation due to the cluster gravitational potential.

The lack of any significant redshift dependence in the $\fe$ -- density relation coupled with the significant
changes with redshift seen in the $\fso$ -- density relation suggest that the origin of the MDR and MRR in high density 
regions is most likely determined by two distinct processes: (i) formation of elliptical galaxies at 
the cluster core, which occurred at redshifts significantly greater than unity and (ii) the 
formation of S0 galaxies, which appears, for at least half of the lenticulars, to be
a process that is still underway even at redshifts as low as $z \sim 0.5$. 
Evidence for different formation timescales for elliptical and S0 galaxies comes from other types of studies as well. 
For example, the analysis of the color-magnitude
relation (CMR) in RDCS1252-2927 by Blakeslee \etal (2003) finds that the observed scatter in the CMR for S0 galaxies
in that system is consistent with relatively recent or even on-going star formation activity whereas the CMR for
elliptical galaxies in this $z=1.24$ cluster is compatible with no significant star formation activity for more than a Gyr. In addition,
Poggianti \etal (1999) and Goto, Yagi, Tanaka, \& Okamura (2004) derive shorter
evolutionary timescales for blue to red spiral transformation than for red spiral to red elliptical transformation,
consistent with the hypothesis that transformation of a galaxy's spectral energy distribution typically 
occurs more rapidly than morphological transformation.

A fundamental question that remains is whether or not the MDR and MRR depend on luminosity. It is well established that
intrinsically luminous, red galaxies are almost always elliptical and, hence, at some level limiting the analyses of the MDR to intrinsically
luminous galaxies may produce a different relationship between morphology and projected density than found in a sample containing a
broader range of luminosities. Tanaka \etal (2005) demonstrate that there is indeed a change in the MDR at $z < 0.07$ as a function
of luminosity based on an analysis of $\sim 20,000$ SDSS galaxies. They find that galaxies fainter than $M^{*}_{r}+1$ do appear to have a different
MDR at low densities than brighter galaxies. We have divided our spectroscopic samples
for MS1054-0321 and RXJ0152-1357 into 
four apparent magnitude ranges $\imag \le 22,\ \imag \le 22.5,\ \imag \le 23,\ {\rm and}\ \imag \le 23.5$ 
corresponding to 
$M^* \le M^*_{775} - 0.3,\ M^* \le M^*_{775} + 0.2,\  M^* \le M^*_{775} + 0.7,\ {\rm and}\ M^* \le M^*_{775} + 1.2$. 
We find no significant difference between the MDR derived for these 4 samples.
If the luminosity dependence of the MDR does not exhibit strong evolution, then our current sample does not go deep 
enough at $z = 0.83$ to accurately measure the dependence. 
Our failure to detect any luminosity dependence in the MDR in our $z = 0.83$ spectroscopic sample is thus
not inconsistent with the work done at low redshifts. Furthermore, our survey is not well suited to studying 
densities below 30 galaxies Mpc$^{-2}$, well above the density where the Tanaka \etal (2005) find the strongest 
luminosity-dependent effects. All we can conclude for now is that for $\Sigma > 30$ galaxies Mpc$^{-2}$ and
for $M^* \simless M^*_{775}+1.2$ the $z = 0.83$ MDR is not sensitive to luminosity. The caveat to this conclusion is that
our \imag\ data at $z = 0.83$ samples the rest-frame $B-$band. It may be that galaxies selected according to a redder
rest-frame luminosity will show a different trend.
The underlying goal of studying the MDR as a function of luminosity is, of course, to assess the
dependence of the MDR on galaxy mass. We are assembling NIR photometry for our cluster galaxies to provide more
accurate stellar mass estimates and we will explore trends between morphology, density, \rtwoh, and galaxy mass in a future paper.

\section{Conclusions}\label{conclude}

We have performed deep, multiband observations with the ACS/WFC of 7 clusters with
$0.83 \le z \le 1.27$ to study the morphological composition of the galaxy population 
over 3 decades of local density and out to radii of up to 2 Mpc from the peak of the 
X-ray emission from the ICM. The key results are:
\begin{enumerate}
\item The high sensitivity and angular sampling of the ACS/WFC enable reliable 
visual distinctions to be made between the major morphological galaxy classes 
(E, S0, Sp+Irr) down to $\sim0.25 L/L^*$ in the range $0.8 < z < 1.3$.
\item We confirm the results of Smith \etal (2004) that a morphology -- density relation exists at $z \sim 1$. We also
explicitly confirm the less rapid growth in $\feso$ with increasing density and that the change
in the slope of the MDR is due primarily to changes in the high-density population fraction.
A flattening of the $\feso$ -- density relation with increasing redshift 
can be a consequence of environmentally-driven
transformation of galaxies from late to early-types. Our new results provide direct measurements of the
lenticular population fraction at $z \sim 1$ and we conclude that the observed differences in the 
morphology -- density relation at $z \sim 1$ are due primarily to a deficit of S0 galaxies and an excess of Sp+Irr galaxies
relative to the local galaxy population. The $\fe$ -- density relation does not appear to evolve over the
range $0 < z < 1.3$. A deeper understanding of the implications of these results for models of galaxy and cluster formation
will require further exploration of the dependence of morphological population fractions 
on galaxy mass and an assessment of the frequency of cluster galaxy merger activity.
\item The MDR at $z = 0.83$ 
is not sensitive to the rest-frame $B-$band luminosity for galaxies with luminosities
brighter than $M^*+1.2$ and in regions with $\Sigma > 30$ galaxies Mpc$^{-2}$. Work done
at low redshift suggests, however, that luminosity effects may be more pronounced at fainter
luminosities and in regions of lower density. Our present survey is not well-suited to studying these
regimes.
\item We directly measure the lenticular population fraction and find $\fso = 0.20 \pm 0.12$ 
when we average over densities with $\Sigma \ge 30$ galaxies Mpc$^{-2}$.
The error in $\fso$ includes contributions from both counting statistics ($\pm 0.035$)
and errors in our ability to visually classify lenticular galaxies ($\pm 0.11$).
Our $z \sim 1\ \fso$ value is about a factor of 2 less than the $\fso$ seen in similarly dense environments in
the local universe but is comparable to what is seen at $0.4 \simless z \simless 0.5$. Our 20\%
population fraction of S0 galaxies is higher than almost all of the scenarios proposed by Smith \etal (2004)
to explain the shallower $\feso$ -- density relation -- they predict $\fso < 0.1$. However, a small reduction in the
elliptical population fraction at $z = 0.5$ adopted by Smith \etal (2004) from $\fe = 0.6$ to $\fe = 0.5$,
a value that appears to be in better agreement with the observations, is sufficient to increase their predicted $z = 1$
S0 population fractions to $0.2\pm0.1$, overlapping our measurement. The distribution of ellipticities in
$z > 0.8$ bulge-dominated (E+S0) cluster galaxies 
is inconsistent with an ellipticity distribution that would arise from a sample consisting solely of
elliptical galaxies. In other words, our results suggest that 
rich clusters at $z \sim 1$ most likely have a significant population of lenticular galaxies and, therefore,
a significant percentage of lenticular galaxies could have formed at redshifts $z > 1.3$. 
\item We measure the morphology -- radius relation and find its evolution is consistent with that
seen in the morphology -- density relation:
\begin{enumerate}
\item the bulk of the transition from a $\fsp$ consistent with that in the field environment to its
minimum value occurs within a radius of 0.6\rtwoh\ (which corresponds, on average, to 
densities $>70$ galaxies Mpc$^{-2}$ and to physical scales less than $\sim750$ kpc), 
\item the $z \sim 1$ $\feso$ value, at a given radius, is systematically less than
the low-$z$ $\feso$ for radii less than $\sim\rtwoh$, and
\item the $\fso$ -- radius relation shows the most
significant difference from the current epoch relationship. 
\end{enumerate}
However, elongation of and clumpiness in the galaxy distributions for many of
our $z > 0.8$ clusters makes interpretation of the azimuthally-averaged morphology -- radius relation
more difficult (and perhaps less meaningful) than trends between morphology and local density. 
\item We find that the bulge-dominated galaxy population fraction, $\feso$, is mildly correlated
with the bolometric X-ray luminosity of the cluster. Clusters with high X-ray luminosities have higher $\feso$ values
within \rtwoh\ than clusters with lower X-ray luminosities. In the present sample, the trend is significant at the $\simless 3\sigma$ level .
A correlation between $\feso$ and bolometric X-ray luminosity can arise as a consequence either of
environmentally-driven transformation processes or initial conditions. However, we do not find significant correlations
between $\feso$ and the X-ray temperature or cluster velocity dispersion.
A definitive study of the relation between galaxy
population fraction and cluster X-ray properties will require the analysis of large homogeneously selected cluster samples. 

\end{enumerate}

\section*{Acknowledgments}

ACS was developed under NASA contract NAS5-32865 and this research has 
been supported by NASA grant NAG5-7697. The STScI is operated by AURA Inc., 
under NASA contract NAS5-26555. We are grateful to Ken Anderson, Jon McCann, 
Sharon Busching, Alex Framarini, Sharon Barkhouser, and Terry Allen for 
their invaluable contributions to the ACS project at JHU. We wish to thank the referee, Alan Dressler,
for his insightful comments. 

\clearpage
\begin{table}
\scriptsize
\caption{Summary of Spectroscopic and HST Observations}
\label{tab_obssum}
\begin{center}
\begin{tabular}{lrrr|c|l|c}
\tableline
\tableline
                     &                   &             &                                    & {Mosaic Area}   &                                            &                  \\
{Cluster}     & {Redshift} & N$_z$ & N$_{z,CL,ACS}$   & (sq.arcmin)        & {Filter (Exp. Time in ksec)}   & HST ID \\
\tableline
MS\,1054$-$0321   & 0.831 & 327          & 143          &  35.5 & \Vmag\ (2.0), \imag\ (4.0), \zmag\ (4.0)  & 9290, 9919\\ 
RXJ\,0152$-$1357  & 0.837 & 123          & 93            & 36.5  & \rmag\ (4.8), \imag\ (4.8), \zmag\ (4.8)   & 9290 \\
CL\,1604$+$4304   & 0.900 & 107          &  20           & 12.0 & \Vmag\ (4.8), \Imag\ (4.8)                      & 9919 \\
CL\,1604$+$4321   & 0.921 & 130          &  31           & 12.0 & \Vmag\ (4.8), \Imag\ (4.8)                      & 9919 \\
RDCS\,0910$+$5422  & 1.101 & $\sim 10$ & $\sim 10$ & 12.2 & \imag\ (6.8), \zmag\ (11.4)                      & 9919 \\
RDCS\,1252$-$2927& 1.235 & 180          &  31           & 32.7 & \imag\ (7.2), \zmag\ (12.0)                     & 9290 \\
RXJ\,0849$+$4452  & 1.266 & 90            & 16            & 33.7 & \imag\ (7.3), \zmag\ (12.2)                     & 9919 \\
\tableline
\end{tabular}
\end{center}
\end{table}

\begin{table}
\scriptsize
\caption{Summary of Cluster X-ray and Kinematic Data}
\label{tab_xray}
\begin{center}
\begin{tabular}{lrccccc|c|c}
\tableline
\tableline
              &   & $\lx$                                  &  $\lxbol$               &   $\tempx$                   &   $\sigma$  & \rtwoh &  X-ray Data & Vel. Disp. \\
{Cluster} &   & (10$^{44}\ h^{-2}_{70}$ erg s$^{-1}$) & (10$^{44}\ h^{-2}_{70}$ erg s$^{-1}$) & (keV)   & (km s$^{-1}$) & ($h^{-1}_{70}$ Mpc) & Reference & Reference \\
\tableline
MS\,1054$-$0321  &~~& 7.78 $\pm$ 0.4      & 16.43 $\pm$ 0.8    & $8.0_{-1.5}^{+2.5}$    & 1153$\pm80$& 1.79 &1, 2 & 10\\ 
RXJ\,0152$-$1357 &~~& 5.74 $\pm$ 0.6      & 18.57 $\pm$ 1.9    & $6.46_{-1.2}^{+1.7}$    & 919$\pm168$& 1.42 & 2, 3 & 11\\
CL\,1604$+$4304  &~~& 0.86 $\pm$ 0.13    &   2.01 $\pm$ 0.3   & $2.51_{-0.69}^{+1.05}$ & 989$^{+98}_{-76}$ & 1.48 & 4 & 12\\
CL\,1604$+$4321   &~~& $<0.7$                 & \nodata                & \nodata                           & 649$^{+59}_{-46}$ & 0.95 &   &  12\\
RDCS\,0910$+$5422 &~~& 0.78 $\pm$ 0.09    &  2.14 $\pm$ 0.3    & $7.20_{-1.4}^{+2.2}$    & \nodata                     & 0.99$^{a}$ & 5 &  \\ 
RDCS\,1252$-$2927 &~~& 1.90 $\pm$ 0.3       &  6.60 $\pm$ 1.1   & $6.50 \pm 0.5$            & 760$^{+117}_{-69}$ & 0.94 & 6, 7 & 13 \\
RXJ\,0849$+$4452 &~~& 1.41 $\pm$ 0.3      &  2.85 $\pm$ 0.6   & $5.80_{-1.7}^{+2.8}$    & \nodata                     & 0.91$^{a}$ & 8, 9 &  \\
\tableline
\end{tabular}
\tablenotetext{a}{r$_{200}$ based on assumed $\sigma = 750$ km s$^{-1}$}
\tablenotetext{1}{Gioia \etal 2004; Donahue 2004, priv. comm.}
\tablenotetext{2}{Romer \etal 2000} 
\tablenotetext{3}{Della Ceca \etal 2000}
\tablenotetext{4}{Lubin \etal 2004}
\tablenotetext{5}{Stanford \etal 2002}
\tablenotetext{6}{Rosati \etal 2004}
\tablenotetext{7}{Lombardi \etal 2005}
\tablenotetext{8}{Rosati \etal 1999}
\tablenotetext{9}{Stanford \etal 2001}
\tablenotetext{10}{Gioia \etal 2004}
\tablenotetext{11}{Demarco \etal 2004}
\tablenotetext{12}{Gal \& Lubin 2004}
\tablenotetext{13}{Demarco \etal 2004a}
\end{center}
\end{table}
\clearpage

\begin{table}
\scriptsize
\caption{Population Fractions as a Function of Projected Density}
\label{tab_mdr_values}
\begin{center}
\begin{tabular}{l|r|c|c|c|c|c|}
\tableline
\tableline
Cluster or &                   & log$_{10}(\Sigma)$   &            &         &        &   \\
Sample &N$(\Sigma)$& (Gals Mpc$^{-2}$)   & $\feso$ & $\fe$ & $\fso$ & $\fsp$ \\
\tableline
$z \sim 1$ composite &   136 ~&~  1.30 ~&~  0.45 $\pm$  0.11 ~&~  0.32 $\pm$  0.14 ~&~  0.13 $\pm$  0.14 ~&~  0.56 $\pm$  0.11 \\ 
                     &   220 ~&~  1.60 ~&~  0.35 $\pm$  0.09 ~&~  0.22 $\pm$  0.13 ~&~  0.13 $\pm$  0.13 ~&~  0.65 $\pm$  0.09 \\ 
                     &   237 ~&~  1.90 ~&~  0.53 $\pm$  0.09 ~&~  0.31 $\pm$  0.13 ~&~  0.22 $\pm$  0.13 ~&~  0.47 $\pm$  0.09 \\ 
                     &   138 ~&~  2.20 ~&~  0.62 $\pm$  0.10 ~&~  0.39 $\pm$  0.14 ~&~  0.23 $\pm$  0.14 ~&~  0.38 $\pm$  0.10 \\ 
                     &    87 ~&~  2.50 ~&~  0.71 $\pm$  0.12 ~&~  0.47 $\pm$  0.15 ~&~  0.24 $\pm$  0.15 ~&~  0.29 $\pm$  0.12 \\ 
                     &    24 ~&~  2.80 ~&~  0.65 $\pm$  0.21 ~&~  0.53 $\pm$  0.23 ~&~  0.12 $\pm$  0.23 ~&~  0.35 $\pm$  0.21 \\ 
                     &    14 ~&~  3.10 ~&~  0.87 $\pm$  0.27 ~&~  0.71 $\pm$  0.29 ~&~  0.16 $\pm$  0.29 ~&~  0.12 $\pm$  0.27 \\ 
\tableline
$z \sim 1$ composite & 185 & 1.06 $\pm$ 0.2 & ~0.30 $\pm$ 0.10 ~& ~0.24 $\pm$ 0.13 ~& ~0.06 $\pm$ 0.13 ~& ~0.70 $\pm$ 0.10 \\
                     & 595 & 2.06 $\pm$ 0.2 & ~0.51 $\pm$ 0.07 ~& ~0.31 $\pm$ 0.12 ~& ~0.20 $\pm$ 0.12 ~& ~0.49 $\pm$ 0.07 \\
                     &  38 & 3.06 $\pm$ 0.2 & ~0.73 $\pm$ 0.17 ~& ~0.59 $\pm$ 0.20 ~& ~0.14 $\pm$ 0.20 ~& ~0.27 $\pm$ 0.17 \\
\tableline
$z < 0.6$ composite  & 435 & 2.06 $\pm$ 0.2 & ~0.63 $\pm$ 0.08 ~& ~0.34 $\pm$ 0.12 ~& ~0.29 $\pm$ 0.12 ~& ~0.37 $\pm$ 0.08 \\
                     &  66 & 3.06 $\pm$ 0.2 & ~0.80 $\pm$ 0.14 ~& ~0.50 $\pm$ 0.17 ~& ~0.30 $\pm$ 0.17 ~& ~0.20 $\pm$ 0.14 \\
\tableline 
MS\,1054$-$0321    & 79 & $\ge2.00$ & ~0.80 $\pm$ 0.13 ~& ~0.49 $\pm$ 0.16 ~& ~0.31 $\pm$ 0.16 ~& ~0.20 $\pm$ 0.13 \\
RXJ\,0152$-$1357   & 63 & $\ge2.00$ & ~0.81 $\pm$ 0.14 ~& ~0.56 $\pm$ 0.17 ~& ~0.25 $\pm$ 0.17 ~& ~0.20 $\pm$ 0.14 \\
CL\,1604$+$4304    & 62 & $\ge2.00$ & ~0.53 $\pm$ 0.14 ~& ~0.33 $\pm$ 0.17 ~& ~0.20 $\pm$ 0.17 ~& ~0.47 $\pm$ 0.14 \\
CL\,1604$+$4321    & 86 & $\ge2.00$ & ~0.39 $\pm$ 0.12 ~& ~0.25 $\pm$ 0.15 ~& ~0.14 $\pm$ 0.15 ~& ~0.61 $\pm$ 0.12 \\
RDCS\,0910$+$5422   & 43 & $\ge2.00$ & ~0.50 $\pm$ 0.16 ~& ~0.37 $\pm$ 0.19 ~& ~0.13 $\pm$ 0.19 ~& ~0.50 $\pm$ 0.16 \\
RDCS\,1252$-$2927  & 46 & $\ge2.00$ & ~0.85 $\pm$ 0.16 ~& ~0.62 $\pm$ 0.18 ~& ~0.23 $\pm$ 0.18 ~& ~0.15 $\pm$ 0.16\\
RXJ\,0849$+$4452   &104 & $\ge2.00$ & ~0.35 $\pm$ 0.11 ~& ~0.33 $\pm$ 0.14 ~& ~0.02 $\pm$ 0.14 ~& ~0.65 $\pm$ 0.11 \\
\tableline
\end{tabular}
\tablenotetext{~}{Errors are the quadrature sum of counting statistics and classification uncertainty. See text for details.}
\end{center}
\end{table}

\clearpage
\begin{table}
\scriptsize
\caption{Population Fractions as a Function of r$_{200}$ Radius}
\label{tab_mrr_values}
\begin{center}
\begin{tabular}{l|r|c|c|c|c|c|}
\tableline
\tableline
Cluster or &          &                    &            &         &           &           \\
Sample    &N(r/\rtwoh)  & r/\rtwoh & $\feso$ & $\fe$ & $\fso$ & $\fsp$ \\
\tableline
$z \sim 1$ composite  &    63 ~&~  0.05 ~&~  0.75 $\pm$  0.14 ~&~  0.52 $\pm$  0.17 ~&~  0.23 $\pm$  0.17 ~&~  0.26 $\pm$  0.14 \\ 
                      &    80 ~&~  0.15 ~&~  0.52 $\pm$  0.13 ~&~  0.37 $\pm$  0.16 ~&~  0.15 $\pm$  0.16 ~&~  0.48 $\pm$  0.13 \\ 
                      &    90 ~&~  0.25 ~&~  0.56 $\pm$  0.12 ~&~  0.35 $\pm$  0.15 ~&~  0.21 $\pm$  0.15 ~&~  0.44 $\pm$  0.12 \\ 
                      &    83 ~&~  0.35 ~&~  0.45 $\pm$  0.13 ~&~  0.24 $\pm$  0.16 ~&~  0.21 $\pm$  0.16 ~&~  0.55 $\pm$  0.13 \\ 
                      &    94 ~&~  0.45 ~&~  0.42 $\pm$  0.12 ~&~  0.25 $\pm$  0.15 ~&~  0.17 $\pm$  0.15 ~&~  0.58 $\pm$  0.12 \\ 
                      &    99 ~&~  0.55 ~&~  0.27 $\pm$  0.12 ~&~  0.21 $\pm$  0.15 ~&~  0.06 $\pm$  0.15 ~&~  0.73 $\pm$  0.12 \\ 
                      &   114 ~&~  0.65 ~&~  0.32 $\pm$  0.11 ~&~  0.20 $\pm$  0.15 ~&~  0.12 $\pm$  0.15 ~&~  0.69 $\pm$  0.11 \\ 
                      &   129 ~&~  0.75 ~&~  0.30 $\pm$  0.11 ~&~  0.15 $\pm$  0.14 ~&~  0.15 $\pm$  0.14 ~&~  0.70 $\pm$  0.11 \\ 
                      &    96 ~&~  0.85 ~&~  0.35 $\pm$  0.12 ~&~  0.11 $\pm$  0.15 ~&~  0.24 $\pm$  0.15 ~&~  0.64 $\pm$  0.12 \\ 
                      &    74 ~&~  0.95 ~&~  0.25 $\pm$  0.13 ~&~  0.15 $\pm$  0.16 ~&~  0.10 $\pm$  0.16 ~&~  0.75 $\pm$  0.13 \\ 
                      &    23 ~&~  1.05 ~&~  0.32 $\pm$  0.22 ~&~  0.32 $\pm$  0.24 ~&~  0.00 $\pm$  0.24 ~&~  0.68 $\pm$  0.22 \\ 
                      &    11 ~&~  1.15 ~&~  0.19 $\pm$  0.31 ~&~  0.00 $\pm$  0.32 ~&~  0.19 $\pm$  0.32 ~&~  0.81 $\pm$  0.31 \\ 
\tableline
MS\,1054$-$0321     ~&~ 130 ~&~ $\le 1.0$ ~&~ 0.78 $\pm$  0.11 ~&~  0.46 $\pm$  0.14 ~&~  0.32 $\pm$  0.14 ~&~  0.22 $\pm$  0.11 \\
RXJ\,0152$-$1357    ~&~ 125 ~&~ $\le 1.0$ ~&~ 0.72 $\pm$  0.11 ~&~  0.46 $\pm$  0.14 ~&~  0.26 $\pm$  0.14 ~&~  0.27 $\pm$  0.11 \\
CL\,1604$+$4304     ~&~ 124 ~&~ $\le 1.0$ ~&~ 0.57 $\pm$  0.11 ~&~  0.40 $\pm$  0.14 ~&~  0.17 $\pm$  0.14 ~&~  0.42 $\pm$  0.11 \\
CL\,1604$+$4321     ~&~ 150 ~&~ $\le 1.0$ ~&~ 0.39 $\pm$  0.10 ~&~  0.27 $\pm$  0.14 ~&~  0.12 $\pm$  0.14 ~&~  0.60 $\pm$  0.10 \\
RDCS\,0910$+$5422   ~&~ 146 ~&~ $\le 1.0$ ~&~ 0.38 $\pm$  0.10 ~&~  0.32 $\pm$  0.14 ~&~  0.06 $\pm$  0.14 ~&~  0.61 $\pm$  0.10 \\
RDCS\,1252$-$2927 ~&~   67 ~&~ $\le 1.0$ ~&~ 0.68 $\pm$  0.14 ~&~  0.47 $\pm$  0.16 ~&~  0.21 $\pm$  0.16 ~&~  0.32 $\pm$  0.14 \\
RXJ\,0848$+$4452   ~&~ 214 ~&~ $\le 1.0$ ~&~ 0.30 $\pm$  0.09 ~&~  0.29 $\pm$  0.13 ~&~  0.01 $\pm$  0.13 ~&~  0.70 $\pm$  0.09 \\
\tableline
\end{tabular}
\tablenotetext{~}{Errors are the quadrature sum of counting statistics and classification uncertainty. See text for details.}
\end{center}
\end{table}

\appendix

\section{Comparing Projected Density Estimation Techniques}\label{app_compden}

We demonstrate here that the Nearest Neighbor (NN) and Friends-of-Friends (FoF) algorithms for
estimating local projected density yield consistent results. We also show that it is possible to construct
a composite MDR or MRR using a combination of galaxy samples selected using spectroscopic redshifts, photometric
redshifts, and samples with only statistically subtracted background corrections providing each of the samples are confined
to specific density ranges.

The FoF algorithm provides an alternative to the NN algorithm as a means to measure the morphological
composition as a function of density.
The FoF algorithm, first developed by Huchra \& Geller (1982), 
has been used widely thereafter for automated detection of overdensities in galaxy catalogs. 
This approach has the advantage that it is not tied to a 
specific choice of $N$ nearest neighbors although it is not the method typically used by other 
workers in the analysis of the MDR.
The FoF algorithm locates groups of galaxies by identifying all the neighboring systems that lie 
within a given ``percolation" length of a given galaxy.  
Each neighbor is then searched for galaxies within the same percolation length. 
This linking process is continued until no additional members can be identified. 
The percolation length and the local projected density are related as
$l_{perc} \propto 1/\sqrt{\Sigma}$. For our comparison,
we use percolation lengths that correspond to overdensities in the range $15 < \Sigma < 1500$
galaxies Mpc$^{-2}$. At each overdensity, the morphological populations within all
identified groups are tallied, weighted by the inverse of the selection function. 
The difference between the cumulative E, S0, and Sp+Irr 
population at two overdensities $\Sigma$ and $\Sigma - \delta\Sigma$
then defines the morphological populations at projected density $\Sigma$. As with the nearest neighbor algorithm,
we correct the FoF densities to correspond to our fiducial luminosity limit 
using equation~\ref{eq_fcorr}.  

\begin{figure*}[ht]
\plotone{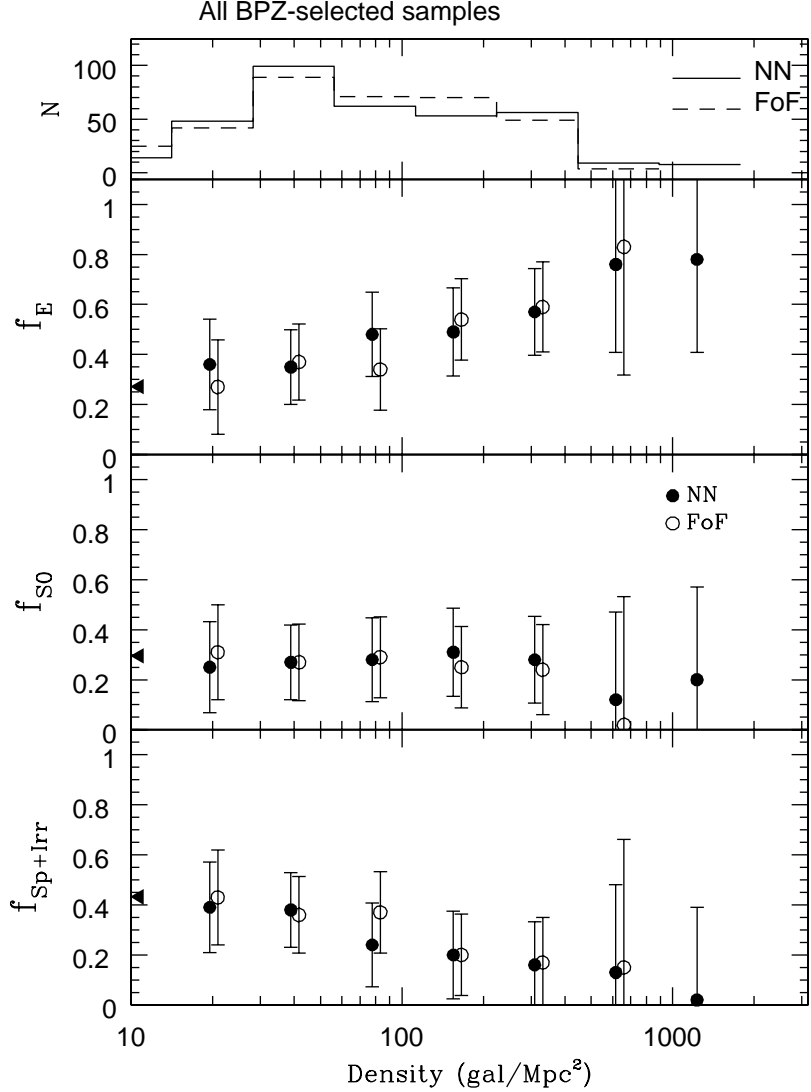}
\caption[]{
The composite morphology -- density relation for the clusters MS1054-0321, RXJ0152-1357, and RDCS1252-2927. 
These results are based on the photo-z selected samples that include galaxies with $z_{ph}$ 
values in the range $|\overline z_{cl} - z_{ph}| / (1 + z_{cl}) \le 2 \sigma_{ph}$.
Results from the $N$th nearest-neighbor (NN) and friends-of-friends (FoF) density estimators are 
both shown here. The density values have been offset by a small amount from one another for clarity.
The total number of galaxies in each density bin is shown in the top panel of the plot.
The low-density population fractions from the SDSS (Goto \etal 2003a) are denoted by the triangles along the y-axis.  }
\label{fig_mdr_bpz}
\end{figure*}

Figure~\ref{fig_mdr_bpz} shows a comparison between the 
NN and FoF density estimators for the clusters for which we have reliable
photometric redshifts (MS1054-0321, RXJ0152-1357, and RDCS1252-2927). The errorbars in all these 
figures include the uncertainties from counting statistics and
from the intrinsic classification uncertainties quantified in section~\ref{morph} 
(\eg $\pm0.06$ in $\feso$ and $\fsp$; $\pm0.11$ in $\fso$ and $\fe$). 
The two density estimators produce consistent results over the full range of densities analyzed.

\subsection{Using Composite Cluster Samples}

\begin{figure*}[ht]
\plotone{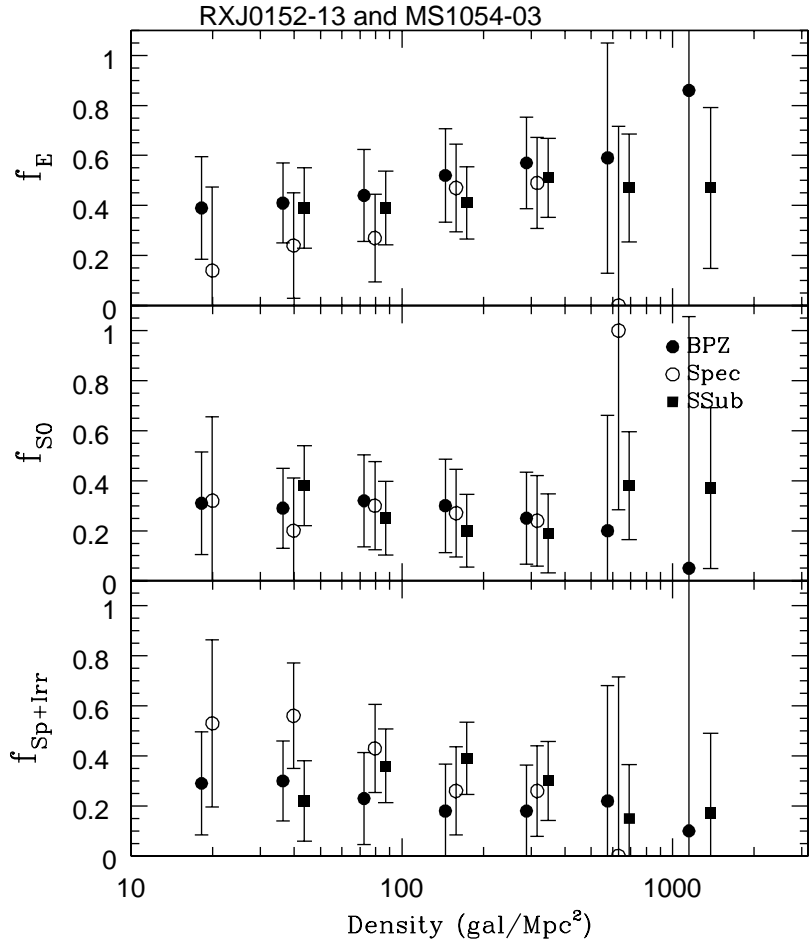}
\caption[]{
The dependence of MDR on sample selection is demonstrated for MS1054-0321 and RXJ0152-1357.
We show the population fractions as a function of projected density based
on the samples of confirmed spectroscopic members (Spec), the samples of photo-z selected members (BPZ), and
on the full flux-limited samples with statistically subtracted background corrections applied (SSub). 
The latter results are only shown
when $\Sigma > 40$ galaxies Mpc$^{-2}$. Errorbars include the uncertainties in counting statistics
and morphological classification.}
\label{fig_compare}
\end{figure*}

Because not all clusters in our sample have sufficient data to compute the local density
based on either spectroscopic or on photometric redshift values,
it is also important to establish that our results are not very sensitive to the details used
to derive the MDR (\eg whether the sample is selected based on photometric redshift value,
spectroscopic redshift value, or a flux-limited sample with statistical background subtraction). 
If we can demonstrate this then we can 
construct a composite MDR by combining for
each individual cluster the sample that yields the most reliable available estimate
(\eg the samples with the least contamination from foreground/background objects).
The clusters RXJ0152-1357 and MS1054-0321 allow us to perform this 
test as they have sufficient information to independently 
generate the MDR from spectroscopic data, from 
photometric redshift data, and from flux-limited samples with a statistically-subtracted background
correction applied. The results are shown in Figure~\ref{fig_compare}. For densities greater than
80 galaxies Mpc$^{-2}$ there are no significant systematic differences in any of the derived
population fractions as a function of density. Below 80 galaxies Mpc$^{-2}$, the statistically
subtracted background corrected density and population fraction estimates become less reliable as fluctuations in
the background galaxy surface density become comparable with the projected density (see
discussion in section~\ref{density}).
The spectroscopic samples exhibit a lower elliptical galaxy fraction at densities below about
50 galaxies Mpc$^{-2}$ but the differences lie within the $1\sigma$ uncertainties. 
The spectroscopic data do not provide good sampling of the very highest densities 
($\Sigma > 1000$ galaxies Mpc$^{-2}$) 
because of the number of slit masks used.
However, in this regime, the photo-z and/or statistically subtracted background results
are very reliable. We conclude that the MDR and MRR derived from a composite sample
will be reliable if spectroscopically selected samples are limited to regions with $\Sigma \le 1000$ galaxies Mpc$^{-2}$
and samples with statistically subtracted background corrections are limited to regions with $\Sigma > 80$ galaxies Mpc$^{-2}$.

\section{Computing Morphological Population Fractions}\label{app_popfrac}

\begin{figure*}
\plotone{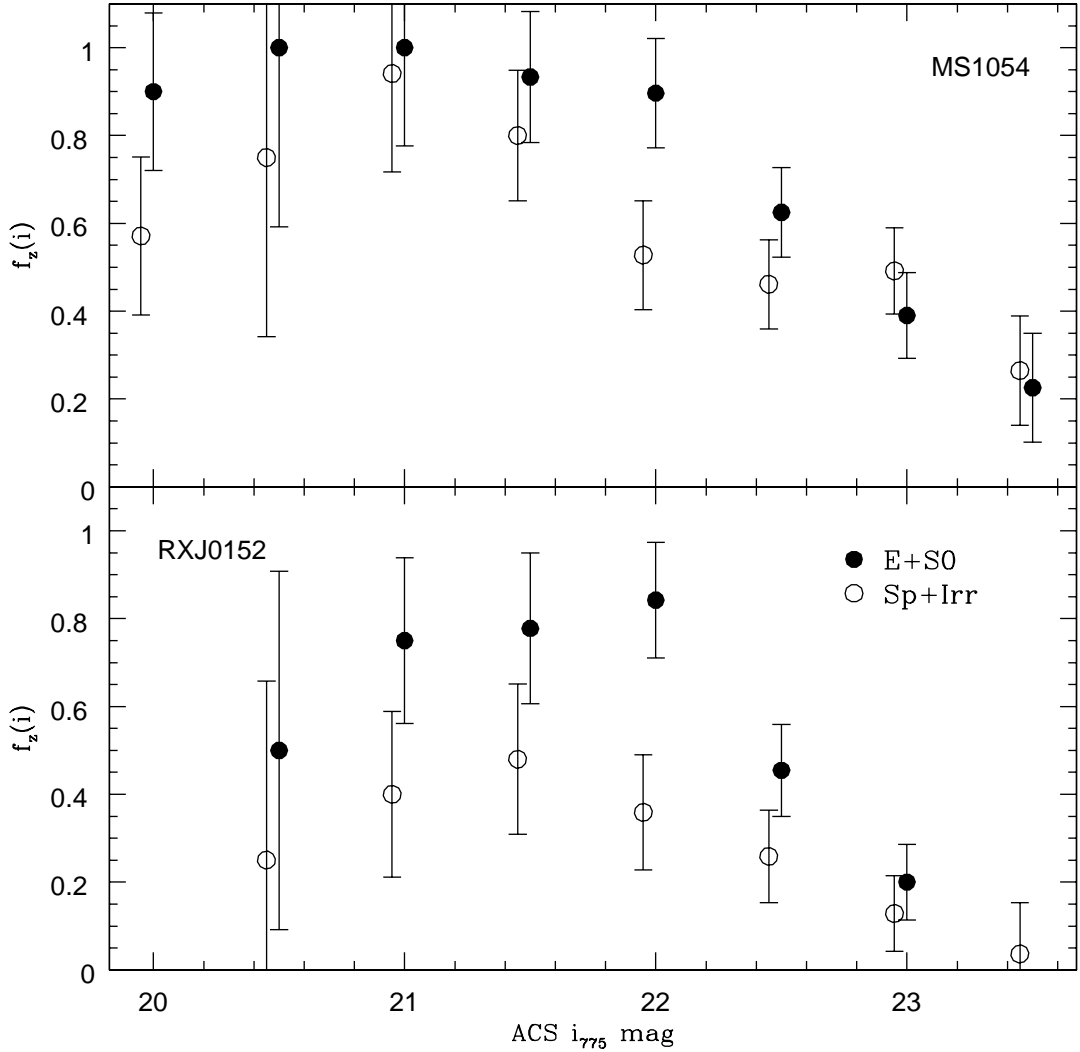}
\caption[]{
The redshift selection functions for E+S0 and  Sp+Irr galaxies vs. \imag\ magnitude 
for MS1054-0321 and RXJ0152-1357. The larger difference between E+S0 and Sp+Irr selection efficiency in 
RXJ0152-1357 is caused by the color selection criterion used in target selection for that cluster.}
\label{fig_zsel}
\end{figure*}

Our morphological population fractions are corrected for both 
contamination and incompleteness in a manner that depends
on the sample being analyzed. For samples using only confirmed 
spectroscopic redshifts, the only correction applied is
the weighting by the inverse of the redshift selection function 
(cf. equation~\ref{eq_density}). 
The redshift selection function is computed empirically by
measuring the ratio of the number of redshifts acquired
to the total number of galaxies in magnitude bins 0.5 mag wide.
The redshift selection function is dependent on the galaxy magnitude 
and to a lesser degree on the galaxy color. 
The latter dependence translates to a dependence on
morphology for those galaxies in the cluster. By measuring the
redshift completeness as a function of color and magnitude, we
can make reasonable corrections for the observational selection
effects. The clusters for which we have a sufficient number of redshifts
to compute the MDR solely from a spectroscopic sample
are MS1054-0321 (with 143 confirmed members, all within the boundaries 
of our ACS mosaic) and RXJ0152-1357 
(with 102 confirmed members, 93 of which lie within the boundaries of our 
ACS mosaic). The redshift selection functions
for these two clusters is shown as a function of \imag\ magnitude and 
morphological type in Figure~\ref{fig_zsel}. 

For photometric redshift selected samples (MS1054-0321, RXJ0152-1357, and RDCS1252-2927), 
we correct the population fractions as 
follows:
\begin{equation}
N_{T}^{corr} = N_{T}^{uncor} + N_{T}^{missed} - N_{T}^{contam} \label{eq_ncorr}
\end{equation}
where $N_{T}^{uncor}$ is the observed number of galaxies of morphological type $T$,
$N_{T}^{missed}$ is an estimate of the number of cluster members of type $T$ that have been
excluded by our photo-z selection criteria, and $N_{T}^{contam}$ is an estimate of the number
of non-cluster members of type $T$ that have been included in the observed count. The
number of missed members is computed as follows:
\begin{equation}
N_{T}^{missed} = N_{total} \times f_{tot,miss} \times f_{miss,T}
\end{equation}
where $N_{total}$ is the total number of galaxies counted at a given density,
$f_{tot,miss}$ is the fraction of cluster members excluded by our photo-z selection
limits, and $f_{miss,T}$ is the fraction of those excluded galaxies that have
morphological type $T$. The value of $f_{tot,miss}$ is estimated by counting the
number of spectroscopically confirmed members that have photo-z's outside our
photo-z selection limits. 
For MS1054-0321 and RXJ0152-1357, $f_{tot,miss}  = 0.38\ {\rm and}\ 0.20$, respectively. 
For RDCS1252-2927, $f_{tot,miss}  = 0.04$. The morphological dependence of the number 
of true cluster galaxies missed by our photo-z selection is weak for MS1054-0321 and RDCS1252-2927 but
is more significant for RXJ0152-1357 (see Figure~\ref{fig_zsel}). 
For MS1054-0321 and RDCS1252-2927, we set $f_{miss,T} $ equal to the initial
uncorrected population fraction for galaxies of type $T$ at the given density. For RXJ0152-1357 we 
assign $f_{miss,T} $ using a morphological distribution that is slightly more heavily weighted to late-type
galaxies at a given density to compensate for the spectroscopic selection bias in this cluster.
Fortunately, the results are not strongly dependent on the precise prescriptions for setting
the $f_{miss,T} $ values. 
The number of non-cluster members contaminating our photo-z selected counts is computed 
as follows:
\begin{equation}
N_{T}^{contam} = N_{total} \times f_{tot,contam} \times f_{contam,T}
\end{equation}
where $f_{tot,contam}$ is the fraction of non-cluster members included by
our photo-z selection limits and $f_{contam,T}$ is the fraction of these that
have morphological type $T$. The value of $f_{tot,contam}$ is estimated by
counting the number of spectroscopically confirmed non-cluster members
that have photo-z's lying within our photo-z selection limits. 
For MS1054-0321 and RXJ0152-1357, $f_{tot,contam}  = 0.09\ {\rm and}\ 0.12$, respectively. 
For RDCS1252-2927, $f_{tot,contam}  = 0.33$. These non-cluster galaxies have a morphological 
distribution that is representative of the field galaxy population and 
we thus set $f_{contam,E} = 0.10,\ f_{contam,S0} = 0.25,
\ {\rm and}\ f_{contam,Sp+Irr} = 0.65$. Note that because we only have $\sim 35$ spectroscopic
redshifts for RDCS1252-2927, the incompleteness and contamination estimates have
larger uncertainties than those in MS1054-0321 or RXJ0152-1357. 

For population fractions derived from magnitude limited samples 
using only statistically-subtracted background counts (\eg as in RDCS0910+5422), the
morphological distribution of the background is assumed to be representative of the field
galaxy population and we thus divide the expected background counts into E, S0, and Sp+Irr
components assuming the same fractions given above: $f_E = 0.10,\ f_{S0} = 0.25,\ {\rm and}
f_{Sp+Irr}=0.65$.


\begin{thebibliography}{}
\bibitem[Abadi \etal(1999)]{ab99} Abadi, M. G., Moore, B., Bower, R. G. 1999, \mnras, 308, 947
\bibitem[Abraham, Valdes, Yee, \& van den Bergh(1994)]{1994ApJ...432...75A}      Abraham, R.~G., Valdes, F., Yee, H.~K.~C., \& van den Bergh, S.\ 1994, \apj, 432, 75 
\bibitem[Abraham \& van den Bergh(2001)]{2001Sci...293.1273A} Abraham,      R.~G.~\& van den Bergh, S.\ 2001, Science, 293, 1273 
\bibitem[Bartko \etal(2005)]{fb05} Bartko, F. \etal 2005, in preparation.
\bibitem[Bekki(1998)]{bekki} Bekki, K. 1998, \apj, 502, 133
\bibitem[Bekki \etal(2002)]{bek02} Bekki, K., Couch, W. J., Shioya, Y. 2002, \apj, 577, 651
\bibitem[Ben\'{i}tez(2000)]{bpz} Ben\'{i}tez, N. 2000, \apj, 536, 57
\bibitem[Ben\'{i}tez \etal(2004)]{b04} Ben\'{i}tez, N., \etal  2004, \apjs, 150, 1
\bibitem[Benson \etal(2001)]{ben01} Benson, A. J., Frenk, C. S., Baugh, C. M., Cole, S., Lacey, C. G. 2001,
   \mnras, 327, 1041
\bibitem[Bertin \& Arnouts(1996)]{1996A&AS..117..393B} Bertin, E.~\& Arnouts, S.\ 1996, \aaps, 117, 393 
\bibitem[Blakeslee et al.(2003a)]{2003adass..12..257B} Blakeslee, J.~P.,      Anderson, K.~R., Meurer, G.~R., Ben{\'{\i}}tez, N., \& Magee, D.\ 2003a, ASP      Conf.~Ser.~295: Astronomical Data Analysis Software and Systems XII, 12, 257 
\bibitem[Blakeslee et al.(2003b)]{2003ApJ...596L.143B} Blakeslee, J.~P., et al.\ 2003b, \apjl, 596, L143 
\bibitem[Bouwens, Broadhurst, \& Silk(1998)]{1998ApJ...506..557B} Bouwens,       R., Broadhurst, T., \& Silk, J.\ 1998, \apj, 506, 557 
\bibitem[Bunker \etal(2000)]{} Bunker, A., Spinrad, H., Stern, D., Thompson, R., 
   Moustakas, L., Davis, M., Dey, A. 2000, astro-ph/0004348
\bibitem[Byrd \& Valtonen(1990)]{bv90} Byrd, G., \& Valtonen, M. 1990, \apj, 350, 89
\bibitem[Carlberg, Yee, \& Ellingson(1997)]{1997ApJ...478..462C} Carlberg,     R.~G., Yee, H.~K.~C., \& Ellingson, E.\ 1997, \apj, 478, 462 
\bibitem[Conselice, Bershady, \& Jangren(2000)]{2000ApJ...529..886C}      Conselice, C.~J., Bershady, M.~A., \& Jangren, A.\ 2000, \apj, 529, 886 
\bibitem[De Propris \etal(2004)]{2004MNRAS.351..125D} De Propris, R., \etal 2004, \mnras, 351, 125
\bibitem[Della Ceca et al.(2000)]{2000A&A...353..498D} Della Ceca, R.,    Scaramella, R., Gioia, I.~M., Rosati, P., Fiore, F., \& Squires, G.\ 2000, \aap, 353, 498
\bibitem[Demarco \etal(2004)]{dem04} Demarco, R., Rosati, P., \etal 2004, \aap, submitted
\bibitem[Demarco et al.(2004a)]{2004cgpc.sympE..10D} Demarco, R., Rosati,     P., Lidman, C., Nonino, M., Mainieri, V., Stanford, A., Holden, B., \& Eisenhardt, P.\ 2004a, 
    {\it Carnegie Observatories Astrophysics Series, Vol 3:
    Clusters of Galaxies: Probes of Cosmological Structure and Galaxy Evolution}, 
     eds. J. Mulchaey, A. Dressler, and A. Oemler.
\bibitem[Demarco \etal(2005)]{dem05} Demarco, R. \etal 2005, in preparation
\bibitem[Donahue et al.(1998)]{1998ApJ...502..550D} Donahue, M., Voit,    G.~M., Gioia, I., Lupino, G., Hughes, J.~P., \& Stocke, J.~T.\ 1998, \apj, 502, 550 
\bibitem[Dressler(1980)]{d80} Dressler, A. 1980, \apj, 236, 351 [D80]
\bibitem[Dressler \etal(1997)]{d97} Dressler, A. \etal 1997, \apj, 490, 577 [D97]
\bibitem[Ettori et al.(2004)]{2004MNRAS.tmp..322E} Ettori, S., et al.\    2004, \mnras, 322, in press
\bibitem[Fabricant \etal(2000)]{fab} Fabricant, D., Franx, M., van Dokkum, P. 2000, \apj, 539, 577
\bibitem[Fairley et al.(2002)]{2002MNRAS.330..755F} Fairley, B.~W., Jones,    L.~R., Wake, D.~A., Collins, C.~A., Burke, D.~J., Nichol, R.~C., \& Romer,    A.~K.\ 2002, \mnras, 330, 755 
\bibitem[Farouki \& Shapiro(1980)]{fs80} Farouki, R., \& Shapiro, S. L. 1980, \apj, 241, 928
\bibitem[Fasano et al.(2000)]{2000ApJ...542..673F} Fasano, G., Poggianti, 
    B.~M., Couch, W.~J., Bettoni, D., Kj{\ae}rgaard, P., \& Moles, M.\ 2000, \apj, 542, 673
\bibitem[Ferguson \etal(2004)]{ferg04} Ferguson, H. C., \etal 2004, \apj, 600, 107
\bibitem[Finn, Zaritsky, \& McCarthy(2004)]{2004ApJ...604..141F} Finn,     R.~A., Zaritsky, D., \& McCarthy, D.~W.\ 2004, \apj, 604, 141
\bibitem[Ford \etal(2003)]{acs} Ford, H. C., Clampin, M., \etal 2003, Proc. SPIE, 4854, 81
\bibitem[Fujita(1998)]{fuj98} Fujita, Y. 1998, \apj, 509, 587
\bibitem[Fujita \& Nagashima(1999)]{fn99} Fujita, Y., \& Nagashima, M. 1999, \apj, 516, 619
\bibitem[Gal \& Lubin(2004)]{2004ApJ...607L...1G} Gal, R.~R.~\& Lubin,     L.~M.\ 2004, \apjl, 607, L1 
\bibitem[Gioia et al.(2004)]{2004A&A...419..517G} Gioia, I.~M., Braito, V., Branchesi, M., Della Ceca, R., 
    Maccacaro, T., \& Tran, K.-V.\ 2004, \aap, 419, 517 
\bibitem[Goto \etal(2003a)]{goto03} Goto, T., Yamauchi, C., Fujita, Y., Okamura, S., Sekiguchi, M., 
   Smail, I., Bernardi, M., Gomez, P. 2003a, \mnras, 346, 601
\bibitem[Goto et al.(2003b)]{2003PASJ...55..739G} Goto, T., et al.\ 2003b, \pasj, 55, 739
\bibitem[Goto, Yagi, Tanaka, \& Okamura(2004)]{2004MNRAS.348..515G} Goto,      T., Yagi, M., Tanaka, M., \& Okamura, S.\ 2004, \mnras, 348, 515 
\bibitem[Goto \etal(2004)]{goto04} Goto, T., \etal 2004, \apj, submitted
\bibitem[Gunn \& Gott(1972)]{gg72} Gunn, J. E., \& Gott, R. 1972, \apj, 176, 1
\bibitem[Heavens \etal(2004)]{heav} Heavens, A., Panter, B., Jimenez, R., Dunlop, J. 2004, Nature, 428, 625
\bibitem[Homeier \etal(2004)]{hom04} Homeier, N., \etal 2005, \apj, in press
\bibitem[Huchra \& Geller(1982)]{hg82} Huchra, J. P., \& Geller, M. J. 1982, \apj, 257, 423
\bibitem[Icke(1985)]{icke} Icke, V. 1985, \aap, 144, 115
\bibitem[Kauffmann(1995)]{kauf95} Kauffmann, G. 1995, \mnras, 274, 161
\bibitem[Kent(1981)]{k81} Kent, S. M. 1981, \apj, 245, 805
\bibitem[Kodama et al.(2001)]{2001ApJ...562L...9K} Kodama, T., Smail, I.,     Nakata, F., Okamura, S., \& Bower, R.~G.\ 2001, \apjl, 562, L9 
\bibitem[Kodama \& Smail(2001)]{2001MNRAS.326..637K} Kodama, T.~\& Smail,    I.\ 2001, \mnras, 326, 637 
\bibitem[Larson, Tinsley, \& Caldwell(1980)]{ltc} Larson, R. B., Tinsley, B. M., Caldwell, C. N 1980, \apj, 237, 692
\bibitem[Lavery \& Henry(1988)]{lh88} Lavery, R. J., \& Henry, J. P. 1988, \apj, 330, 596
\bibitem[Lombardi \etal(2005)]{lomb} Lombardi, M., Rosati, P., Blakeslee, J. P., Ettori, S., Demarco, R., Ford, H. C.,
Illingworth, G. D., Clampin, M., Hartig, G. F., Ben\'{i}tez, N., Broadhurst, T. J., Franx, M., Jee, M. J., Postman, M.,
\& White, R. L. 2005, \apj, in press.
\bibitem[Lubin, Mulchaey, \& Postman(2004)]{2004ApJ...601L...9L} Lubin,     L.~M., Mulchaey, J.~S., \& Postman, M.\ 2004, \apjl, 601, L9 
\bibitem[Makino \& Hut(1997)]{mh97} Makino, J., \& Hut, P. 1997, \apj, 481, 83
\bibitem[Mamon(1992)]{mam92} Mamon, G. A. 1992, \apjl, 401, L3
\bibitem[Margoniner, de Carvalho, Gal, \& Djorgovski(2001)]{2001ApJ...548L.143M} Margoniner, V.~E., de Carvalho,      R.~R., Gal, R.~R., \& Djorgovski, S.~G.\ 2001, \apjl, 548, L143 
\bibitem[Moore \etal(1996)]{m96} Moore, B., Katz, N., Lake, G., Dressler, A., Oemler, A. 1996, Nature, 379, 613
\bibitem[Moore \etal(1999)]{m99} Moore, B., Lake, G., Quinn, T., Stadel, J. 1999, \mnras, 304, 465
\bibitem[Moss \& Whittle(2000)]{2000MNRAS.317..667M} Moss, C.~\& Whittle,     M.\ 2000, \mnras, 317, 667 
\bibitem[Mulchaey \& Zabludoff(1998)]{mz98} Mulchaey, J., \& Zabludoff, A. I. 1998, \apj, 496, 73
\bibitem[Okamoto \& Nagashima(2003)]{2003ApJ...587..500O} Okamoto, T.~\& Nagashima, M.\ 2003, \apj, 587, 500 \bibitem[Papovich et al.(2003)]{2003ApJ...598..827P} Papovich, C.,     Giavalisco, M., Dickinson, M., Conselice, C.~J., \& Ferguson,       H.~C.\ 2003, \apj, 598, 827 
\bibitem[Poggianti et al.(1999)]{1999ApJ...518..576P} Poggianti, B.~M., 
Smail, I., Dressler, A., Couch, W.~J., Barger, A.~J., Butcher, H., Ellis, 
R.~S., \& Oemler, A.~J.\ 1999, \apj, 518, 576
\bibitem[Ponman \etal(1996)]{pon} Ponman, T. J., Bourner, P. D. J., Ebeling, H., 
   Bohringer, H. 1996, \mnras, 283, 601
\bibitem[Postman \& Geller(1984)]{PG84} Postman, M. \& Geller, M. J., 1984, \apj, 281, 95 [PG84]
\bibitem[Postman \etal(1998a)]{p98a} Postman, M., Lubin, L. M., Oke, J. B. 1998a, \aj, 116, 560
\bibitem[Postman \etal(1998b)]{p98b} Postman, M., Lauer, T. R., Szapudi, I., Oegerle, W. 1998b, \apj, 506, 33
\bibitem[Postman \etal(2001)]{p01} Postman, M., Lubin, L., Oke, J. B. 2001, \aj, 122, 1125
\bibitem[Quilis \etal(2000)]{q2000} Quilis, V., Moore, B., Bower, R. 2000, Science, 288, 1617
\bibitem[Roche et al.(1998)]{1998MNRAS.293..157R} Roche, N., Ratnatunga,      K., Griffiths, R.~E., Im, M., \& Naim, A.\ 1998, \mnras, 293, 157 
\bibitem[Romer et al.(2000)]{2000ApJS..126..209R} Romer, A.~K., et al.\ 2000, \apjs, 126, 209 
\bibitem[Rosati et al.(1999)]{1999AJ....118...76R} Rosati, P., Stanford,      S.~A., Eisenhardt, P.~R., Elston, R., Spinrad, H., Stern, D., \& Dey, A.\ 1999, \aj, 118, 76 
\bibitem[Rosati et al.(2004)]{2004AJ....127..230R} Rosati, P., et al.\ 2004, \aj, 127, 230 
\bibitem[Schlegel \etal(1998)]{schl} Schlegel, D. J., Finkbeiner, D. P., Davis, M. 1998, 
   \apjs, 500, 525
\bibitem[Smith \etal(2004)]{sm04} Smith, G. P., Treu, T., Ellis, R. S., Moran, S. M., Dressler, A. 2004,
   astro-ph/0403455, \apj, submitted [Sm04]
\bibitem[Stanford \etal(1997)]{stan97} Stanford, S. A., Elston, R., Eisenhardt, P., Spinrad, H.,
    Stern, D., Dey, A. 1997, \aj, 114, 2232
\bibitem[Stanford et al.(2001)]{2001ApJ...552..504S} Stanford, S.~A.,    Holden, B., Rosati, P., Tozzi, P., Borgani, S., Eisenhardt, P.~R., \& Spinrad, H.\ 2001, \apj, 552, 504 
\bibitem[Stanford et al.(2002)]{2002AJ....123..619S} Stanford, S.~A.,    Holden, B., Rosati, P., Eisenhardt, P.~R., Stern, D., Squires, G., \& Spinrad, H.\ 2002, \aj, 123, 619 
\bibitem[Tanaka \etal(2005)]{tan05} Tanaka, M., Goto, T., Okamura, S., Shimasaku, K., Brinkman, J., 2005, \aj, in press.
\bibitem[Toft et al.(2004)]{2004A&A...422...29T} Toft, S., Mainieri, V., 
   Rosati, P., Lidman, C., Demarco, R., Nonino, M., \& Stanford, S.~A.\ 2004, \aap, 422, 29
\bibitem[Tran \etal(2005)]{tran} Tran, K., Magee, D., Franx, M., Illingworth, G. D., 
   Kelson, D., van Dokkum, P. 2005, in preparation.
\bibitem[Treu et al.(2003)]{2003ApJ...591...53T} Treu, T., Ellis, R.~S.,    Kneib, J., Dressler, A., Smail, I., Czoske, O., Oemler, A., \& Natarajan,    P.\ 2003, \apj, 591, 53 
\bibitem[Trujillo et al.(2004)]{2004ApJ...604..521T} Trujillo, I., et al.\ 2004, \apj, 604, 521 
\bibitem[Tully \& Shaya(1984)]{1984ApJ...281...31T} Tully, R.~B.~\& Shaya,    E.~J.\ 1984, \apj, 281, 31 \bibitem[Valluri(1993)]{val93} Valluri, M. 1993, \apj, 408, 57
\bibitem[van Dokkum \etal(2000)]{vd2000}van Dokkum, P. G., Franx, M., Fabricant, D., 
   Illingworth, G. D., Kelson, D. D. 2000, \apj, 541, 95
\bibitem[Whitmore \& Gilmore(1991)]{wg91} Whitmore, B. C., \& Gilmore, D. M. 1991, \apj, 367, 64
\bibitem[Windhorst \etal(2002)]{wind02} Windhorst, R., \etal  2002, \apjs, 143, 113
\bibitem[Xue \& Wu(2000)]{2000ApJ...538...65X} Xue, Y.~\& Wu, X.\ 2000, \apj, 538, 65 
\bibitem[Zabludoff \& Mulchaey(1998)]{zm98} Zabludoff, A. I., \& Mulchaey, J. S. 1998, \apj, 496, 39
\end{thebibliography}
\end{document}